\documentclass[10pt,twocolumn,twoside]{IEEEtran}

\usepackage[ruled,vlined]{algorithm2e}
\usepackage{amssymb}
\usepackage{bm}
\usepackage{cite}
\usepackage{graphicx}
\usepackage{mathtools}
\usepackage{multicol}
\usepackage{multirow}
\usepackage{subfigure}
\usepackage{tcolorbox}
\usepackage{theorem}
\usepackage{tikz}
\usepackage{times,amsmath,epsfig}
\usepackage{url}
\usepackage{xcolor}
\usepackage{comment}

\usetikzlibrary{shapes,arrows}
\input{mysymbol.sty}

\newtheorem{myassumption}{\bf Assumption}
\newtheorem{myremark}{\bf Remark}

\title{SLoG-Net: Algorithm Unrolling for\\ Source Localization on Graphs}

\author{\IEEEauthorblockN{Chang Ye, \textit{Student Member, IEEE,} and Gonzalo Mateos, \textit{Senior Member, IEEE}}
\thanks{Work in this paper is supported in part by the Center of Excellence in Data Science, an Empire State Development-designated Center of Excellence. Part of the results in this paper appeared at the \textit{2022 EUSIPCO Conference}~\cite{chang2022eusipco}. \emph{(Corresponding author: Gonzalo Mateos)}.

Chang Ye and Gonzalo Mateos are with the Department of Electrical and Computer Engineering, University of Rochester, Rochester, NY 14627 USA (e-mail: cye7@ur.rochester.edu; gmateosb@ece.rochester.edu).}}

\begin{document}
\maketitle

\begin{abstract}%
We present a novel model-based deep learning solution for the inverse problem of localizing sources of network diffusion. Starting from first graph signal processing (GSP) principles, we show that the problem reduces to joint (blind) estimation of the forward diffusion filter and a sparse input signal that encodes the source locations. Despite the bilinear nature of the observations in said blind deconvolution task, by requiring invertibility of the diffusion filter we are able to formulate a convex optimization problem and solve it using the alternating-direction method of multipliers (ADMM). We then unroll and truncate the novel ADMM iterations to arrive at a parameterized neural network architecture for Source Localization on Graphs (SLoG-Net), that we train in an end-to-end fashion using labeled data. This supervised learning approach offers several advantages such as interpretability, parameter efficiency, and controllable complexity during inference. Our reproducible numerical experiments corroborate that SLoG-Net exhibits performance on par with the iterative ADMM baseline, but with markedly faster inference times and without needing to manually tune step-size or penalty parameters. Overall, our approach combines the best of both worlds by incorporating the inductive biases of a GSP model-based solution within a data-driven, trainable deep learning architecture for blind deconvolution of graph signals.
\end{abstract}



%
\section{Introduction}\label{sec:intro}

\IEEEPARstart{W}{e} study the inverse problem of localizing sources of network diffusion when the forward model is unknown, also referred to as \emph{blind} graph filter identification~\cite{segarra2017blindid,chang2024exact,chang2024perturbation}. In this problem, we observe $P$ graph signals $\{\bby_i\}_{i=1}^{P}$ that we model as outputs of some diffusion graph filter, i.e., a polynomial in the graph-shift operator of a known graph~\cite{sandryhaila2013discrete,ortega2018gsp,gama2020spmag,isufi2024gf}. The goal is to jointly identify the filter coefficients $\bbh$ and the input signals $\{\bbx_i\}_{i=1}^P$ that generated the network observations. We assume that the input signals are sparse, implying only a few source nodes inject a signal that spreads through the network~\cite{segarra2017blindid}. This problem can be viewed as a form of blind deconvolution in graph domains and has applications in a variety of fields, including sensor-based environmental monitoring, opinion formation in social networks, neural signal processing, epidemiology, or disinformation campaigns. For prior related blind deconvolution approaches of (non-graph) signals, see e.g.,~\cite{ahmed2014blind,levin2011understanding,wang2016blind}.

\subsection{Proposed approach in context}\label{Ss:Related}

To solve this inverse problem in a supervised setting, we propose a novel data-driven machine learning model that blends graph signal processing (GSP) principles with deep learning (DL). Our algorithm unrolling~\cite{monga2021spmag} approach leverages the interpretability and parameter efficiency of model-based GSP methods, while also utilizing the power of DL to achieve satisfactory recovery performance and controllable complexity during inference; see also~\cite{chen2021graphunroll,nagahama2022tsp} for related ideas applied to graph signal denoising and~\cite{shrivastava2019glad,pu2021learning,max2023gdn,wasserman2024bnn} for network topology inference.

Different from most early attempts to localizing sources on networks (e.g.,~\cite{zhang2016towards,pinto2012locating,sefer2016diffusion}), here we leverage the GSP toolbox~\cite{ortega2018gsp} inspired by~\cite{pena2016source,segarra2017blindid,chang2018eusipco}. The idea in~\cite{segarra2017blindid} is to cast the (bilinear) blind graph filter identification task as a \emph{linear} inverse problem in the ``lifted'' rank-one, row-sparse matrix $\bbx \bbh^\top$; see also~\cite{ahmed2014blind,LingBiConvexCS} for seminal blind deconvolution work via convex programming. While the rank and sparsity minimization algorithms in \cite{segarra2017blindid,david_blind_sp} can successfully recover sparse inputs along with low-order graph filters (even from a single observation, i.e., with $P=1$), reliance on matrix lifting can hinder applicability to large graphs. Beyond this computational consideration, the overarching assumption of~\cite{segarra2017blindid} is that the inputs $\{\bbx_i \}_{i=1}^P$ share a common support when $P>1$. Moreover, iterative solvers in~\cite{segarra2017blindid,david_blind_sp,chang2018eusipco} require carefully tuning step-sizes, as well as carrying out costly grid searches to select regularization parameters in the inverse problems. Instead, the algorithm unrolling-based DL model of this paper learns these parameters (and others) in a end-to-end fashion. 

Other works adopt probabilistic models of network diffusion, and resulting maximum-likelihood source estimators can only be optimal for particular (e.g., tree) graphs~\cite{pinto2012locating}, or rendered scalable under restrictive dependency assumptions~\cite{feizi2016network}. Relative to~\cite{pinto2012locating,pena2016source,hu2016localizing}, the proposed framework can accommodate signals defined on general undirected graphs and relies on a \emph{convex} estimator of the sparse sources of diffusion, which here we favorably exploit to design a DL architecture as well as to generate training examples. 


\subsection{Contributions and paper outline}\label{Ss:Outline}

In this context, our starting point is the blind graph filter identification formulation in~\cite{chang2024exact}. After reviewing the necessary GSP background, in Section \ref{sec:preliminaries} we reexamine and state the problem in the novel supervised learning setting dealt with here. The \emph{model-based} approach in~\cite{chang2024exact} imposes a mild requirement on invertibility of the graph filter, which facilitates a convex reformulation for the multi-signal case with arbitrary supports (Section \ref{S:blind_ID}); see also~\cite{wang2016blind} for a time-domain precursor that inspired our line of work in graph settings. While~\cite{chang2024exact} focused on fundamental exact recovery and noise stability guarantees (see also~\cite{victor2024icassp,chang2024perturbation} for robustness to graph perturbations), here we shift gears to algorithmic issues and first develop a novel solver based on the alternating-directions method of multipliers (ADMM)~\cite[Ch. 3.4.4]{Bertsekas_Book_Distr}. The ADMM algorithm in Section \ref{Ss:ADMM} is of independent interest as an effective model-based solution to the problem of source localization on graphs (we reiterate that algorithms were only tangentially treated in e.g.,~\cite{segarra2017blindid,chang2018eusipco,chang2024exact}). However, the ADMM's main upshot here is in leveraging its primal and dual variable updates as the blueprint for (sub-)layers of a trainable parametric DL model with prescribed depth. This way we seek to overcome the burden of manually tuning step-size and penalty parameters, plus the time as well as computational overhead that comes with running hundreds or thousands of iterations to attain convergence each time a new problem instance is presented.

To this end, in Section \ref{S:SLoG-Net} we unroll and truncate the ADMM iterations~\cite{monga2021spmag,yang2020admmunroll,nagahama2022tsp}, to arrive at a parameterized nonlinear DL architecture for \textbf{S}ource \textbf{L}ocalization \textbf{o}n \textbf{G}raphs (SLoG-Net), that we train in an end-to-end fashion using labeled data. This way we leverage inductive biases of a GSP model-based solution in a data-driven trainable deep network, which is interpretable, parameter efficient, and offers controllable complexity during inference~\cite{monga2021spmag}. To increase the model's expressive power we explore several customizations to the vanilla SLoG-Net architecture, such as different parameters across layers and learned linear constraints on the inverse filter response. Experiments with both simulated and real network data in Section \ref{S:simulation} demonstrate that SLoG-Net achieves performance on par with the iterative ADMM baseline, while achieving significant post-training speedups. We also show that the model refinements are indeed effective and that SLoG-Net is robust to noise corrupting the observations.
All in all, our findings show promise for blind deconvolution on graphs, while they also support the broader prospect of adopting algorithm unrolling as a versatile data-driven tool to tackle network inverse problems; see also~\cite{shrivastava2019glad,chen2021graphunroll,pu2021learning,nagahama2022tsp,max2023gdn}. Conclusions are laid out in Section~\ref{S:conclusion}, with a discussion of potential follow-up work in this space. Some non-essential algorithm construction steps, mathematical arguments, and DL model implementation details are deferred to the appendices. In support of current reproducible research practices, we share the code used to generate the results reported in this paper.

Relative to the conference precursor~\cite{chang2022eusipco}, in this journal paper we offer a markedly expanded presentation (including extended discussions, schematic diagrams, and appendices) along with full-blown technical details. Noteworthy additions include: (i) SLoG-Net architectural refinements such as learnable constraints and decoupled parameters across layers; (ii) efficient matrix inversion schemes for the model-based ADMM algorithm and the SLoG-Net filter sub-layer; (iii) computational complexity analyses; and (iv) a comprehensive and reproducible performance evaluation protocol. The latter offers comparisons with the iterative ADMM as well as graph neural network (GNN) baselines~\cite{gama2020spmag,wang2022invertible}; a study of inverse filter recovery and source localization performance as a function of the type of random graph ensemble and the number of nodes $N$; robustness to observation noise and the number of observations $P$; as well as real data experiments using a version of the Digg 2009 dataset~\cite{hogg2012social}.

\noindent\emph{Notation:} The entries of a matrix $\mathbf{X}$ and a (column) vector $\mathbf{x}$ are denoted by $X_{ij}$ and $x_i$, respectively. Sets are represented by calligraphic capital letters and $[\bbX]_{\ccalI}$ denotes a submatrix of $\bbX$ formed by selecting the entries of $\bbX$ indexed by $\ccalI$. The notation $^\top$ stands for transpose; $\mathbf{0}_N$ and $\mathbf{1}_N$ refer to the all-zero and all-one vectors of length $N$. The $N\times N$ identity matrix is denoted by $\bbI_N$. For a vector $\bbx$, $\diag(\mathbf{x})$ is a diagonal matrix whose $i$th diagonal entry is $x_i$. 
The operators $\circ$ and $\odot$ stand for the Hadamard (elementwise)  and Khatri-Rao (columnwise Kronecker) matrix products. 

\section{Preliminaries and Problem Statement}\label{sec:preliminaries}

We start by introducing the basic graph theoretical background required to formally state the inverse problem of localizing sources of network diffusion. Let $G(\ccalV,\bbA)$ denote a weighted and undirected network graph, where $\ccalV=\{1,\ldots,N\}$ is the set of vertices and $\bbA \in \reals_+^{N \times N}$ is the symmetric adjacency matrix. Entry $A_{ij}=A_{ji}\geq 0$ denotes the edge weight between nodes $i$ and $j$. \vspace{2pt}

\noindent \textbf{Graph signals and shift operators.} A graph signal $\bbx: \ccalV  \mapsto \reals^N$ is an $N$-dimensional vector, where entry $x_i$ represents the signal value at node $i \in \ccalV$; see~\cite{ortega2018gsp} for examples in sensor networks, social media, transportation systems, or network neuroscience. As a general algebraic descriptor of network structure relating the entries of $\bbx$, one can define a \textit{graph-shift operator} $\bbS \in \reals^{N \times N}$ which is a matrix with the same sparsity pattern as $G$ \cite{sandryhaila2013discrete}. Accordingly, $\bbS$ can be viewed as a local, meaning one-hop, diffusion (or aggregation) operator acting on graph signals. See~\cite{gavili2017shift,ortega2018gsp} for typical choices including normalized variations of adjacency and Laplacian matrices. Next, we introduce more general operators -- graph filters -- that linearly combine multi-hop aggregations of graph signals obtained via self compositions of $\bbS$. Our particular focus will be on simple generative mechanisms behind network diffusion. 

\subsection{Graph filter models of linear network diffusion}\label{ssec:graph_filters}

Consider a graph signal $\bby$ that is supported on a graph $G$ with shift operator $\bbS$, and is generated from an input state $\bbx$ through \emph{linear network diffusion}. Formally, we can write
\begin{align}\label{eqn_diffusion}
\bby = a_0 \prod_{l=1}^{\infty} (\bbI_N-a_l \bbS) \bbx
= \sum _{l=0}^{\infty} \bar{a}_l \bbS^l \bbx,
\end{align}
where $\bbS$ captures one-hop localized interactions among network nodes, and each successive application of the shift in \eqref{eqn_diffusion} diffuses $\bbx$ over $G$. The signal mapping in \eqref{eqn_diffusion} encompasses several existing models, including heat diffusion, consensus, and the classic DeGroot model of opinion dynamics~\cite{DeGrootConsensus}.

Despite the infinite degree of the polynomial expressions in \eqref{eqn_diffusion}, the Cayley-Hamilton theorem ensures that they are equivalent to polynomials of degree upper bounded by $N$~\cite[pp. 109-110]{horn2013matrixanalysis}. Defining the vector of coefficients $\bbh:=[h_0,\ldots,h_{L-1}]^\top$ and the (convolutional) graph filter 
\begin{equation}\label{e:graph_filter_def}
\mathbf{H}:=h_0\bbI_N+h_1\bbS+\ldots+h_{L-1}\bbS^{L-1}=\sum_{l=0}^{L-1}h_l \mathbf{S}^l,
\end{equation}
the signal model in \eqref{eqn_diffusion} can be rewritten as $\bby = \bbH \bbx$ for some specific $\bbh$ and $L\leq N$. As formalized in the ensuing section, in this paper we adopt $\bby = \bbH \bbx$ as a forward model for the measurements $\bby$. We want to recover $\bbx$ when $\bbh$ is also unknown. For an up to date and comprehensive survey of graph filters; the interested reader is referred to~\cite{isufi2024gf}. \vspace{2pt}

\noindent \textbf{Frequency domain representation.} Since $\bbS$ is symmetric, it is diagonalizable as $\bbS=\bbV\bbLambda\bbV^\top$, with $\bbLambda=\textrm{diag}(\lambda_1,\ldots,\lambda_N)$ collecting the eigenvalues. This spectral decomposition of $\bbS$ is used in GSP to represent graph filters and signals in the graph frequency domain.
Specifically, let us use the eigenvalues of $\bbS$ to define the Vandermonde matrix  $\bbPsi_L\in \reals^{N\times L}$, where $\Psi_{ij}:=\lambda_i^{j-1}$.
The frequency representations of a signal $\bbx$ and filter $\bbh$ are defined as $\tbx:=\bbV^\top\bbx$ and $\tbh:=\bbPsi_L\bbh$, respectively. The former is by definition of the graph Fourier transform (GFT)~\cite{DSP_freq_analysis,ortega2018gsp} and the latter follows since the filter output $\bby\!=\!\bbH\bbx$ in the frequency domain can be written as
\begin{equation} \label{e:freq_response}
	\tby=\diag\big(\bbPsi_L\bbh\big)\bbV^\top \bbx=\diag\big(\tbh\big)\tbx=\tbh\circ \tbx.
\end{equation} 
This identity is analogous to the convolution theorem for temporal signals, where we find $\tby$ is given by the elementwise product $(\circ)$ of $\tbx$ and the filter's frequency response $\tbh$.

\subsection{Problem statement}\label{ssec:prob_statement}

\begin{figure}[t]
	\centering    
	{\includegraphics[width=\linewidth]{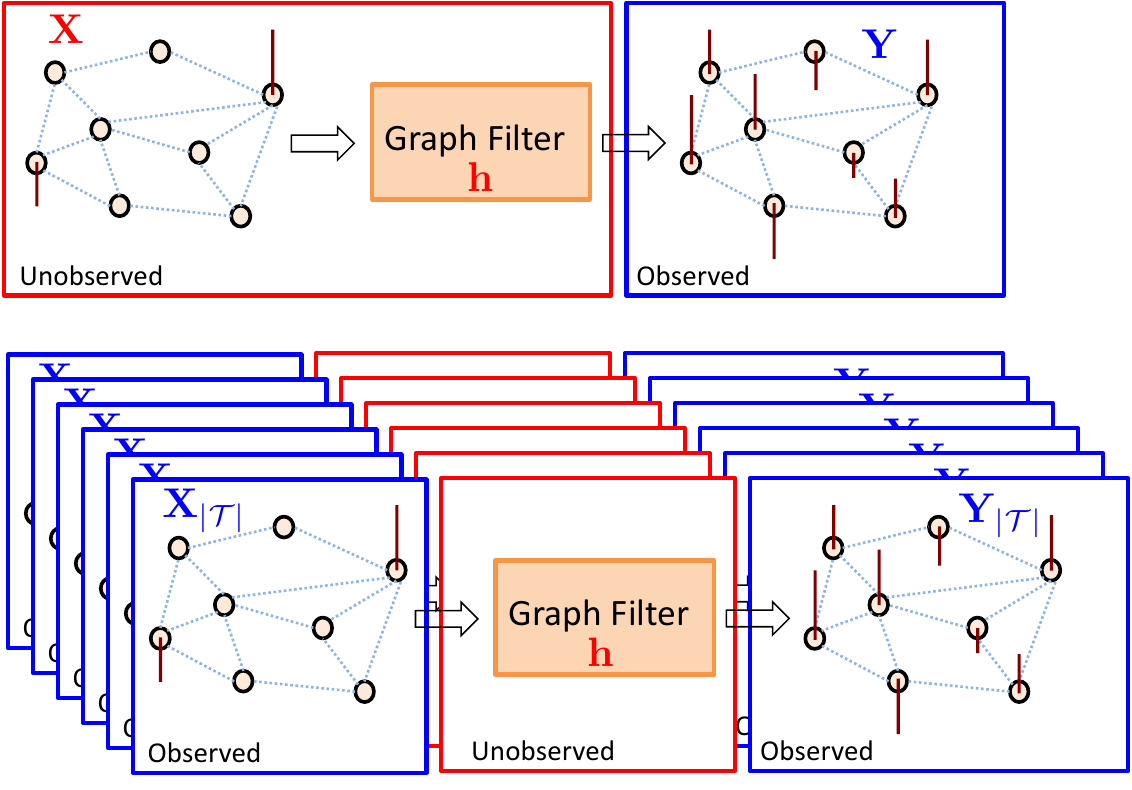}}
 \caption{(top) Model-based source localization on graphs as a blind deconvolution task. Given $P$ graph signals in $\bbY$ modeled as the output of a diffusion graph filter, the goal is to recover the filter coefficients $\bbh$, and the input signals $\bbX$ that are assumed to be sparse. Blue panels indicate what is given and red panels represent unobserved quantities. (bottom) When formulated as a supervised learning problem, we rely on a training set $\ccalT:=\{\bbX_i,\bbY_i\}_{i=1}^{|\ccalT|}$ to learn the parameters $\bbTheta$ of the model $\hbX=\Phi(\bbY;\bbTheta)$. During training, what is observed and what is not differs from the model-based setting -- and the latter is what we encounter during testing or inference.}
	\label{fig:blindid_setup}
\end{figure}

For a given graph $G$ with shift operator $\bbS$, consider a diffusion filter $\bbH=\sum_{l=0}^{L-1}h_l \mathbf{S}^l$ whose coefficients $\bbh$ are unknown. Like the graph topology, the filter order $L\leq N$ is assumed to be given. Suppose we observe $P$ diffused signals that we arrange in a matrix $\bbY=[\bby_1,\ldots,\bby_P] \in \reals^{N \times P}$, where $\bbY = \bbH \bbX$ and latent inputs $\bbX = [\bbx_1,\ldots,\bbx_P]\in \reals^{N \times P}$. We make the following assumption on the input signals.
\begin{myassumption}[Source sparsity]\label{as:sparsity}
Sources $\bbX\in \reals^{N \times P}$ are \emph{sparse} with at most $S\ll N$ non-zero entries per column.
\end{myassumption}
In this context, source localization amounts to jointly estimating sparse $\bbX$ and the filter coefficients $\bbh$ up to scaling and (possibly) permutation ambiguities~\cite{chang2018eusipco}; see also Fig. \ref{fig:blindid_setup} (top) for a depiction of this blind deconvolution task first studied in~\cite{segarra2017blindid}. Assumption \ref{as:sparsity} is well justified when the signals in $\bbY$ represent diffused versions of a \textit{few} localized sources in $G$, here indexed by $\textrm{supp}(\bbX):=\{(i,j) \mid X_{ij} \neq 0 \}$. Moreover, without such structural constraints the problem is ill-posed, because the number of unknowns $NP + L$ in $\{\bbX,\bbh\}$ exceeds the $NP$  observations  in $\bbY$. 

All in all, using \eqref{e:freq_response} the diffused source localization task can be stated as a feasibility problem of the form
\begin{equation} \label{e:blind_feasibility}
	\text{find } \{ \bbX,\bbh \} \:\: \text{s. to }\: \bbY = \bbV\diag\big(\bbPsi_L\bbh\big)\bbV^\top\bbX, \: \| \bbX \|_{0} \leq PS,
\end{equation}
where the $\ell_0$-(pseudo) norm $\| \bbX \|_{0}:=|\textrm{supp}(\bbX)|$ counts the non-zero entries in $\bbX$. In words, we are after the solution to a system of bilinear equations subject to a sparsity constraint in $\bbX$; a hard problem due to the non-convex $\ell_0$-norm as well as the bilinear constraints. To deal with the latter, similar to~\cite{wang2016blind,chang2018eusipco} we will henceforth assume that the filter $\bbH$ is invertible. \vspace{2pt} 

\noindent \textbf{Source localization as a supervised learning problem.} Suppose that $\bbX$ is drawn from some distribution of sparse matrices, say the Bernoulli-Gaussian model for which one can establish \eqref{e:blind_feasibility} is identifiable~\cite[Remark 1]{chang2018eusipco}. Likewise, suppose the filter taps $\bbh$ are drawn from a distribution such that $\bbH$ is invertible with high probability. Then given independent training samples $\ccalT:=\{\bbX_i,\bbY_i\}_{i=1}^{|\ccalT|}$ adhering to \eqref{eqn_diffusion}, our goal in this paper is to learn a judicious parametric mapping that predicts $\hbX=\Phi(\bbY;\bbTheta)$ by minimizing a loss function
\begin{equation}\label{e:loss}
	L(\bbTheta):=\frac{1}{|\ccalT|}\sum_{i\in\ccalT}\ell(\bbX_i,\Phi(\bbY_i;\bbTheta)),
\end{equation}
where $\bbTheta$ are learnable parameters; see Fig. \ref{fig:blindid_setup} (bottom) for a schematic depiction of the training setting. Depending on the application, a training set may be available from historical data, or for instance it may be generated using a simulator of the diffusion process. Alternatively, given observations $\bbY_i$ one can obtain source labels $\bbX_i$ by solving a convex optimization problem as discussed in the ensuing section; see also~\cite{segarra2017blindid,chang2018eusipco}. This way, the perspective is to \emph{learn to approximate the minimizers} of a relaxation to \eqref{e:blind_feasibility}. While admittedly the data-generation and training phases can be time consuming, they are offline and need to be performed once. In turn, the upshot is a fast method during testing or inference.

We will design the deep network $\Phi(\cdot;\bbTheta)$ in Section \ref{S:SLoG-Net}, using iterations of a model-based solution we develop in Section \ref{S:blind_ID} as a layer by layer blueprint. The particular choice of the loss $\ell$ will be discussed in Section \ref{S:simulation}. While not made explicit in our notation, $\Phi(\cdot;\bbTheta)$ makes internal predictions of the diffusion filter from which $\hbX$ is obtained at the output. 
%

\section{Model-based Source Localization on Graphs}\label{S:blind_ID}

Here we review the model-based solution to the blind deconvolution problem proposed in~\cite{chang2018eusipco,chang2024exact}, which relies on a convex relaxation of \eqref{e:blind_feasibility} when the diffusion filter is invertible. Then we develop novel ADMM iterations to solve said relaxation, which we unroll in Section \ref{S:SLoG-Net} to obtain the SLoG-Net model that we train using data by minimizing \eqref{e:loss}.

\subsection{Convex relaxation for invertible graph filters}\label{ssec:convex_relax}

Conditions for the invertibility of a graph filter can be readily obtained in the graph spectral domain. Indeed, the filter's frequency response $\tilde{h}_i$ should not vanish at any of the discrete frequencies $\lambda_i,\: i=1,\ldots,N,$ otherwise the input information in the nulled frequency modes is lost; see \eqref{e:freq_response}. 
\begin{myassumption}[Filter invertibility]\label{as:invertibility}
The graph filter $\bbH=\bbV\diag(\tbh)\bbV^\top$ is invertible, meaning $\tilde{h}_i=\sum_{l=0}^{L-1} h_l \lambda_{i}^l \neq 0$, for all $i=1,\ldots,N$.
\end{myassumption}
Under Assumption \ref{as:invertibility}, one can show that the inverse operator $\bbG := \bbH^{-1}$ is also a polynomial graph filter on $G$, of degree at most $N-1$ \cite[Theorem 4]{sandryhaila2013discrete}. To be more specific, let $\bbg \in \reals^{N}$ be the vector of inverse-filter coefficients, i.e., $\bbH^{-1}:=\bbG= \sum_{l=0}^{N-1} g_l \bbS^l$. Then one can equivalently rewrite the forward model $\bbY=\bbH\bbX$ for the observations as
\begin{equation} \label{e:Filter_G}
	\bbX  = \bbG \bbY  = \bbV \text{diag}(\tbg) \bbV^\top \bbY,
\end{equation}
where $\tilde{\bbg} := \bbPsi_N \bbg \in \reals^N$ is the inverse filter's frequency response and $\bbPsi_N \in \reals^{N \times N}$ is Vandermonde. Naturally, $\bbG=\bbH^{-1}$ implies the condition $\tbg\circ\tbh =\mathbf{1}_N$ on the frequency responses. Leveraging \eqref{e:Filter_G}, one can recast \eqref{e:blind_feasibility} as a \emph{linear} inverse problem
\begin{equation} \label{e:opt_blind_invertible}
	\min_{\{\bbX,\tbg \}}
	\:\| \bbX \|_{0},\:\: \text{s. to }\:
	\bbX = \bbV \text{diag}(\tbg) \bbV^\top \bbY,\:\bbX\neq\mathbf{0}.
\end{equation}

The $\ell_0$ norm in \eqref{e:opt_blind_invertible} makes the problem NP-hard to optimize. Over the last decade or so, convex-relaxation
approaches to tackle sparsity-minimization problems have enjoyed remarkable success, since they often entail no loss of optimality; see also Remark \ref{rem:guarantees}. Accordingly, we instead: (i) seek to minimize the $\ell_1$-norm convex surrogate of the cardinality function, that is $\| \bbX \|_{1,1} = \sum_{i,j}| X_{ij}|$; and (ii) express the filter in the graph spectral domain as in \eqref{e:Filter_G} to obtain the cost function
\begin{align*}
    \| \bbX \|_{1,1} ={}& \| \bbG \bbY \|_{1,1} \\	= {}& \| \bbV \text{diag}(\tilde{\bbg}) \bbV^\top \bbY \|_{1,1}\\
	= {}&\| (\bbY^\top\bbV \odot \bbV) \tilde{\bbg} \|_{1}.
\end{align*}
%
In arriving at the last equality we used that $\| \bbX \|_{1,1}=\| \textrm{vec}[\bbX] \|_{1}$ and invoked properties of the vectorization operator. This suggests solving the convex $\ell_1$-synthesis problem (in this case a linear program), e.g.,~\cite{zhang2016one}, namely
\begin{equation} \label{e:opt_prob_convex}
	\widehat{\tilde{\bbg}}= \argmin_{\tilde{\bbg} \in \reals^{N}}
	\:\| (\bbY^\top\bbV\odot \bbV) \tilde{\bbg} \|_{1},\quad \text{s. to }\:
	\mathbf{1}_N^\top \tilde{\bbg} = 1.
\end{equation}
While the linear constraint in \eqref{e:opt_prob_convex} avoids the trivial solution $\widehat{\tilde{\bbg}} = 0$, it also serves to fix the (arbitrary) scale of the estimated filter. Once the frequency response $\widehat{\tbg}$ of the inverse filter is recovered, the sources can be reconstructed via $\textrm{vec}[\hbX] =  (\bbY^\top\bbV \odot \bbV) \widehat{\tilde{\bbg}}$. Because $\bbS$ and $L$ are known, one can also recover the filter $\bbH$, if so desired.
\begin{myremark}[Recovery and stability guarantees]\label{rem:guarantees}\normalfont
Sufficient conditions were derived in~\cite{chang2024exact}, under which \eqref{e:opt_prob_convex} succeeds in exactly recovering the true inverse filter response with high probability. This result holds for Bernoulli-Gaussian distributed $\bbX$~\cite{li2015unified,wang2016blind}. Stability to additive noise corrupting the observations $\bbY$~\cite{chang2024exact}, or, perturbations in the graph-shift operator eigenbasis $\bbV$ ~\cite{chang2024perturbation}, has also been established.
\end{myremark}

All in all, under Assumption \ref{as:invertibility} one can readily use e.g., an off-the-shelf interior-point method to solve \eqref{e:opt_prob_convex} efficiently~\cite{chang2018eusipco,chang2024exact}. Next, we propose a specialized sparsity-minimization algorithm that exploits the problem's unique structure.

\subsection{ADMM algorithm}\label{Ss:ADMM}

Problem \eqref{e:opt_prob_convex} can be solved using the ADMM. Let $\bbx=\textrm{vec}[\bbX]\in\reals^{NP}$ and denote $\bbZ:= \bbY^\top\bbV\odot \bbV\in \reals^{NP\times N}$. Using variable splitting, problem \eqref{e:opt_prob_convex} can be equivalently written as
\begin{equation} \label{e:opt_prob_convex_admm}
	\min_{\{\bbx,\tbg\}}  \|\bbx\|_1,\:\: \text{s. to }\:
	\bbZ\tilde{\bbg} - \bbx = \mathbf{0}_{NP},\:\:
	\mathbf{1}_{N}^\top\mathbf{\tilde{\bbg}}  = c,
\end{equation}
where $c=1$, but we will henceforth treat it as a generic nonzero constant in case we want to adjust the scale of $\tbg$. Associating dual variables $\bblambda\in \reals^{NP}$ and $\mu\in\reals$ to the equality constraints in \eqref{e:opt_prob_convex_admm}, the augmented Lagrangian function becomes
\begin{align}\label{e:opt_prob_convex_lagrangian}
	\ccalL_\rho(\bbx,\tilde{\bbg},\bblambda,\mu) = {}& 
	\|\bbx \|_1 + \frac{\rho_{\lambda}}{2}\| \bbZ\tilde{\bbg} -\bbx + \bblambda/\rho_{\lambda} \|_2^2\nonumber\\ 
	{}&+ \frac{\rho_{\mu}}{2}(\mathbf{1}_{N}^\top\tilde{\bbg} - c + \mu/\rho_{\mu})^2,
\end{align}
after completing the squares, where $\rho_{\lambda}$ and $\rho_{\mu}$ are non-negative penalty coefficients. Letting $\bbGamma:=\rho_\lambda\bbZ^\top\bbZ+\rho_{\mu}\mathbf{1}_N\mathbf{1}_N^\top$ for notational convenience, then the ADMM~\cite{boyd2011distributed,Bertsekas_Book_Distr} updates are given by ($k=0,1,2,\ldots$ will henceforth denote iterations)
\begin{align} 
\tilde{\bbg}[k+1] = {}&  \bbGamma^{-1}\left[\bbZ^\top(\rho_\lambda\bbx[k]-\bblambda[k]) + (\rho_\mu c - \mu[k])\mathbf{1}_N\right],\label{eq:admm_update rule_g}\\
\bbx[k+1] ={} &  \ccalS_{\rho_\lambda^{-1}}(\bbZ\tilde{\bbg}[k+1]+\bblambda[k]/\rho_\lambda)\label{eq:admm_update rule_x},\\
\bblambda[k+1] ={} &  \bblambda[k] + \rho_\lambda(\bbZ\tilde{\bbg}[k+1] - \bbx[k+1]),\label{eq:admm_update rule_lambda}\\
\mu[k+1]={} & \mu[k] + \rho_\mu(\mathbf{1}_N^\top\tilde{\bbg}[k+1] - c).\label{eq:admm_update rule_mu}
\end{align}
For completeness, \eqref{eq:admm_update rule_g}-\eqref{eq:admm_update rule_mu} are derived in Appendix \ref{app:admm}. The soft-thresholding operator $S_{\rho_\lambda^{-1}}(\cdot)=\textrm{sign}(\cdot)\max(|\cdot|-\rho_\lambda^{-1},0)$ in \eqref{eq:admm_update rule_x} acts component-wise on the entries of its vector argument. The initialization $\{\bbx[0],\bblambda[0],\mu[0]\}$ can be arbitrary, and we typically let the initial conditions be equal to zero. 

Different from the solvers in~\cite{segarra2017blindid,david_blind_sp}, the provably convergent ADMM updates are free of expensive singular-value decompositions per iteration. The inversion of the $N\times N$ matrix $\bbGamma$ is done once, offline, and $\bbGamma^{-1}\bbZ^\top$, $\bbGamma^{-1}\mathbf{1}_N$ are cached to run the iterations. Even more, we show in Appendix \ref{app:diag_structure} that the matrix $\bbZ\bbZ^\top$ is diagonal. Thus, $\bbGamma$ is a rank-one correction of a diagonal matrix, which can be computed efficiently using the matrix inversion lemma; see Appendix \ref{app:mil_rank_one} for the details and associated computational complexity analysis.

In the next section, we unroll the ADMM iterations \eqref{eq:admm_update rule_g}-\eqref{eq:admm_update rule_mu} to arrive at the trainable parametric model $\Phi(\bbY;\bbTheta)$.


\section{Source Localization via Algorithm Unrolling } \label{S:SLoG-Net}

The algorithm unrolling (or deep unfolding) principle was pioneered in~\cite{gregor2010lista} for the problem of sparse coding natural images using overparameterized dictionaries. The technical approach in~\cite{gregor2010lista} was to truncate and map \emph{iterations} of the iterative shrinkage-thresholding algorithm (ISTA)~\cite{beck2009fista} to \emph{layers} in a deep network that can be trained from data. Learnable weights are often optimization step-sizes, regularization and penalty parameters, or other matrices when additional expressive power is needed. One of the greatest DL challenges has been architectural search, and unrolling offers a principled approach to model design by using tested algorithms as architectural templates. The perspective is to \emph{learn to approximate} solutions with substantial computational savings during inference, relative to the optimization algorithm. While the former process entails a forward pass through a feedforward neural network (NN) with few layers, the latter could entail running hundreds (pr thousands) of iterations until convergence. 

Beyond parsimonious signal modeling, there has been a surge in popularity of unrolled deep networks for a wide variety of applications in signal and image processing; see e.g.,~\cite{monga2021spmag} for a recent review. Most relevant to our approach is the unrolling of ADMM iterations for undersampled image reconstruction~\cite{yang2020admmunroll}, and recent advances to learn from graph data~\cite{chen2021graphunroll,pu2021learning,wasserman2024bnn,max2023gdn,shrivastava2019glad,nagahama2022tsp}. However, none of these works has dealt with the source localization task over networks, and in particular via the blind deconvolution formulation in Section \ref{S:blind_ID} (which has broader applicability; see Section \ref{sec:intro}).

\begin{figure*}
	\centering    
	\includegraphics[width=\textwidth]{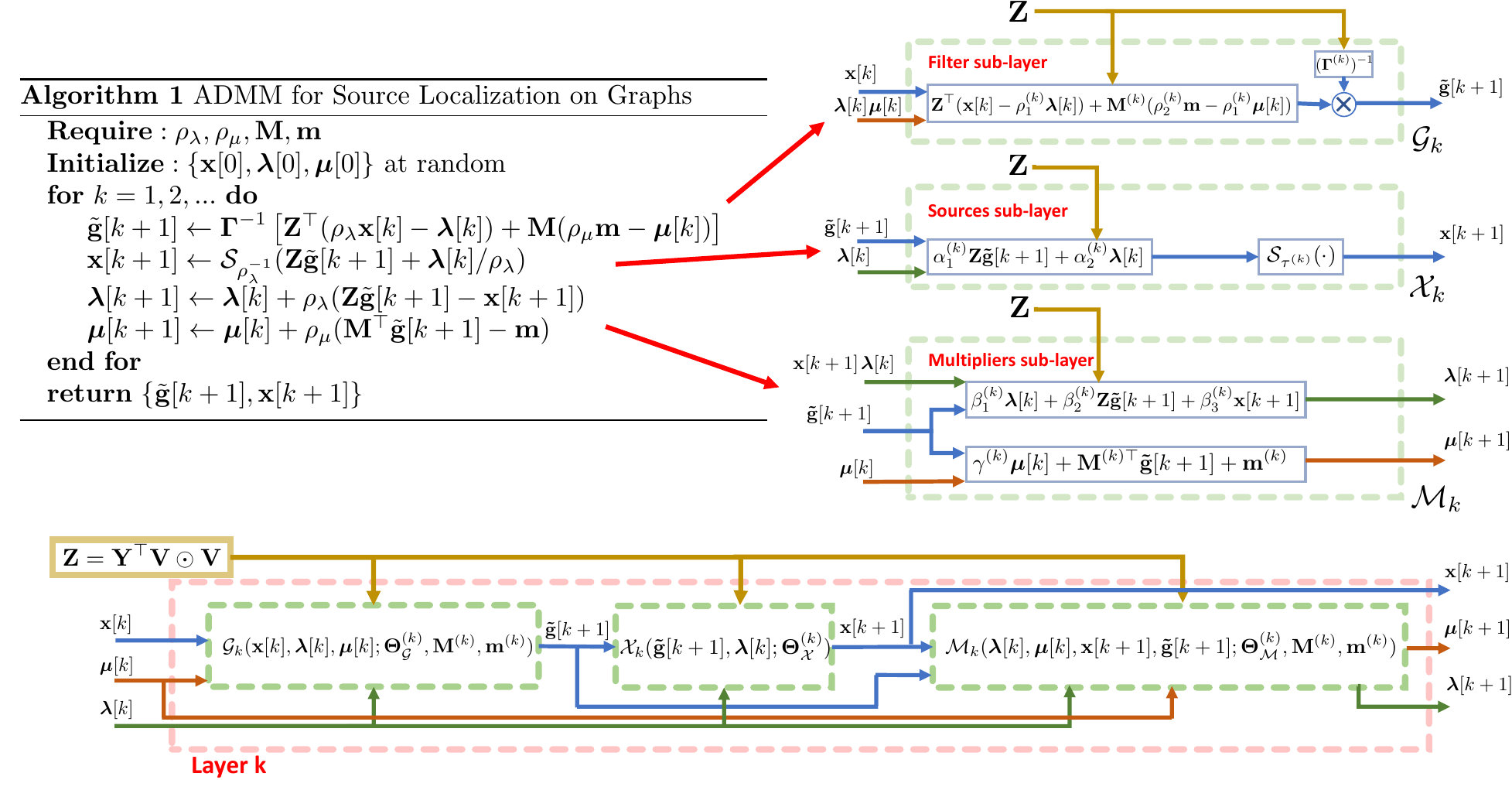}
	\caption{The layerwise structure of SLoG-Net, a DL model $\hbX=\Phi(\bbY;\bbTheta)$ consisting of $K$ layers. Layer $k$ is a composition of three sub-layers: (i) a filter sub-layer $\ccalG_k$; followed by (ii) a sources sub-layer $\ccalX_k$; followed by (iii) a multipliers sub-layer $\ccalM_k$. Each of these sub-layers is a direct mapping of a primal or dual variable update in the ADMM algorithm of Section \ref{Ss:ADMM}, tabulated as Algorithm 1 for convenience. Learnable parameters in $\bbTheta$ are $\bbTheta_\ccalG^{(k)} = \left\{\rho_1^{(k)},\rho_2^{(k)} \right\}$, $\bbTheta_\ccalX^{(k)} = \left\{\alpha_1^{(k)}, \alpha_2^{(k)},\alpha_3^{(k)},\tau^{(k)}\right\}$, $\bbTheta_\ccalM^{(k)} = \left\{\beta_1^{(k)},\beta_2^{(k)},\beta_3^{(k)},\gamma^{(k)}\right\},\bbM^{(k)},\bbm^{(k)}$, $k=1,\ldots,K$.}
	\label{fig:blockdiag}
\end{figure*}

\subsection{SLoG-Net: ADMM as architectural blueprint}\label{Ss:SLoG-Net_Architecture}

We construct the SLoG-Net architecture by unrolling the ADMM iterations \eqref{eq:admm_update rule_g}-\eqref{eq:admm_update rule_mu} into a DL model. To this end, we map individual primal and dual variable updates as sub-layers within a layer; see Fig. \ref{fig:blockdiag} for a schematic depiction of this process. We then compose a prescribed number $K$ of layers to constitute the parametric mapping $\Phi(\bbY;\bbTheta)$. ADMM penalty coefficients $\{\rho_\lambda,\rho_\mu\}$ will be treated as learnable parameters in $\bbTheta$. Observations $\bbY$ are inputs to the NN. Just like in the model-based approach in Section \ref{S:blind_ID}, the architecture leverages graph structure information through the eigenvectors $\bbV$. The $K$-th layer output is used to form the source predictions $\hbX$. \vspace{2pt}

\noindent \textbf{Architectural refinements.} In designing SLoG-Net's sub-layers, we will deviate slightly from a strict ADMM unrolling of \eqref{eq:admm_update rule_g}-\eqref{eq:admm_update rule_mu} in order to enhance overall predictive performance. For instance, in each sub-layer we introduce several additional parameters to increase the model's expressive power. 

In the original formulation \eqref{e:opt_prob_convex}, we included the linear constraint $\mathbf{1}^\top \tilde{\bbg} = 1$ as a simple mechanism to prevent an undesirable all-zero solution and fix the (otherwise arbitrary) scale of the solution. However, this rigid constraint might limit the method's recovery potential in some cases. 
In addition, the scale fixing parameter $c=1$ is no longer needed in the supervised learning setting dealt with here. Scale information will be implicitly conveyed through examples $\ccalT:=\{\bbX_i,\bbY_i\}_{i=1}^{|\ccalT|}$, and can thus be learned during training. These considerations motivate replacing the constraint $\mathbf{1}^\top \tilde{\bbg} = c$ in \eqref{e:opt_prob_convex_admm} with $\bbM^\top \tilde{\bbg} = \bbm$, where $\bbM\in\mathbb{R}^{N\times d}$ and $\bbm\in\mathbb{R}^{d}$ are learnable parameters. The associated modifications to the ADMM algorithm in Section \ref{Ss:ADMM} are straightforward, and we will not spell them out here in the interest of brevity. 

We will also forgo the parameter sharing constraint imposed by the unrolled ADMM iterations. This is standard practice~\cite{monga2021spmag}, and we have experimentally found that a model with different parameters per layer performs better and leads to more stable training. We would be remiss by not mentioning there is a tradeoff, since a NN with coupled parameters offers the flexibility to train with $K_{\textrm{train}}$ layers and then perform inference over a deeper architecture with $K_{\textrm{test}}>K_{\textrm{train}}$ layers. We leave this exploration as future work. Next, we describe the design of each sub-layer in detail. \vspace{2pt}

\noindent \textbf{Filter sub-layer.} This sub-layer $\ccalG_k$ refines the inverse filter coefficient estimate $\tilde{\bbg}[k+1]$ at layer $k$, based on the source estimates $\bbx[k]$ and the dual variables $\{\bblambda[k],\bbmu[k]\}$ from the previous layer. We mimic the $\tilde{\bbg}$ update in \eqref{eq:admm_update rule_g}, and introduce some minor tweaks. To avoid problems with the inversion of $\bbGamma$ in the eventuality $\rho_\lambda=\rho_\mu=0$ during training, we opt for the reparameterization $\rho_1 := 1/\rho_\lambda$, $\rho_2 := \rho_\mu/\rho_\lambda$ and impose non-negativity constraints on both parameters. Moreover, we consider decoupled parameters $\bbTheta_\ccalG^{(k)}:=\big\{\rho_1^{(k)},\rho_2^{(k)}\big\}$ across layers $k=1,\ldots,K$, to increase the network capacity. Finally, the constraint's constant vector $\mathbf{1}_N$ and the scale-normalization constant $c$ are replaced by learnable parameters $\big\{\bbM^{(k)},\bbm^{(k)}\big\}_{k=1}^K$, thus obtaining [cf. \eqref{eq:admm_update rule_g}]
%
%
%
\begin{align}\label{eq:layer_g}
	\tilde{\bbg}[k+1] = {}& \left(\bbZ^\top\bbZ+\rho_2^{(k)}\bbM^{(k)}\bbM^{(k)\top}\right)^{-1}\left[\bbZ^\top\left(\bbx[k]\phantom{\rho_1^{(k)}}\right.\right. \nonumber\\
	&\left.\left.-\rho_1^{(k)}\bblambda[k]\right)+ \bbM^{(k)}\left(\rho_2^{(k)}\bbm^{(k)} - \rho_1^{(k)}\bbmu[k]\right) \right]\nonumber\\
    := {}&\ccalG_k\left(\bbx[k],\bblambda[k],\bbmu[k];\bbTheta_\ccalG^{(k)},\bbM^{(k)},\bbm^{(k)}\right),
\end{align}
where $\rho_1^{(k)},\rho_2^{(k)} \geq 0$, for $k=1,\ldots,K$. 

Recall that $\bbZ:= \bbY^\top\bbV\odot \bbV$, so every time a new data mini-batch $\bbY$ is to be processed one needs to (re)invert matrices $\bbGamma^{(k)}$; see Appendices \ref{app:mil_rank_one} and \ref{app:mil_general} for a computationally-efficient implementation, especially when $d=1$.\vspace{2pt}

\noindent \textbf{Sources sub-layer.} Here we update the source estimates $\bbx[k]$ based on $\tilde{\bbg}[k+1]$ in \eqref{eq:layer_g} and the multiplier $\bblambda[k]$. The sub-layer $\ccalX_k$ imitates \eqref{eq:admm_update rule_lambda}, but instead of a single tunable parameter $\rho_{\lambda}$ we introduce learnable combination weights $\big\{\alpha_1^{(k)},\alpha_2^{(k)}\big\}_{k=1}^{K}$ and thresholds  $\big\{\tau^{(k)}\big\}_{k=1}^{K}$; all collected in $\bbTheta_\ccalX^{(k)}$, for each sub-layer $k=1,\ldots,K$. We propose [cf. \eqref{eq:admm_update rule_x}]
\begin{align} \label{eq:layer_x}
	\bbx[k+1] = {}&\ccalS_{\tau^{(k)}}\left(\alpha_1^{(k)}\bbZ\tbg[k+1]+\alpha_2^{(k)}\bblambda[k]\right)\nonumber\\
    := {}& \ccalX_k\left(\tbg[k+1],\bblambda[k];\bbTheta_\ccalX^{(k)}\right),
\end{align}
where the sparsifying thresholds are naturally constrained as $\tau^{(k)}\geq 0$, for $k=1,\ldots,K$. Notice how \eqref{eq:layer_x} implements a simple linear filter on the sub-layer inputs $\{\tbg[k+1],\bblambda[k1]\}$, followed by a point-wise nonlinear activation, which is reminiscent of vanilla NN layers.\vspace{2pt}

\noindent \textbf{Multipliers sub-layer}. In this simple linear sub-layer $\ccalM_k$, we perform parallel updates of the Lagrange multipliers  $\{\bblambda[k+1],\bbmu[k+1]\}$ by combining $\{\bblambda[k],\bbmu[k]\}$ from layer $k-1$ and the primal variables $\{\tilde{\bbg}[k+1],\bbx[k+1]\}$. Layer-$k$ combination weights $\bbTheta_\ccalM^{(k)}$ are learnable parameters 
$\big\{\beta_1^{(k)},\beta_2^{(k)},\beta_3^{(k)}\big\}_{k=1}^{K}$ and $\big\{\gamma^{(k)}\big\}_{k=1}^{K}$, leading to [cf. \eqref{eq:admm_update rule_lambda}-\eqref{eq:admm_update rule_mu}]
%
%
\begin{align} 
	\bblambda[k] & {}=  \beta_1^{(k)}\bblambda[k-1] + \beta_2^{(k)}\bbZ\tilde{\bbg}[k] + \beta_3^{(k)}\bbx[k], \label{eq:layer_lambda_new}\\
	\bbmu[k] & {}=  \gamma^{(k)}\bbmu[k-1] + \bbM^{(k)\top}\tilde{\bbg}[k]+\bbm^{(k)}. \label{eq:layer_mu_new}
\end{align}

Each SLoG-Net layer is thus a sequential composition of these three sub-layers, in the order we have introduced them: first the filter sub-layer $\ccalG_k$, then the sources $\ccalX_k$, and finally the multipliers sub-layer $\ccalM_k$. Notice how the data embedded in $\bbZ$ is not only fed to the first layer $K=1$, but to all subsequent layers in a way akin to residual neural networks (ResNets).

In closing, we note that the intial states $\{\bbx[0],\bblambda[0],\mu[0]\}$ can be: (i) used as a means to incorporate prior information (especially on the source locations $\bbx$); (ii) randomly initialized as we do in the ensuing experiments; or (iii) learned from data along with $\bbTheta$ as it is customary with recurrent neural networks (RNNs). Going all the way to layer $K$, source predictions are generated as $\hbX=\Phi(\bbY;\bbTheta)=\textrm{unvec}[(\bbY^\top\bbV \odot \bbV)\tilde{\bbg}[K]].$ 

Given a training set $\ccalT:=\{\bbX_i,\bbY_i\}_{i=1}^{|\ccalT|}$ of e.g., syntethic data, or, real signals $\bbY_i$ and source estimates obtained using ADMM, learning is accomplished by using mini-batch stochastic gradient descent to minimize the loss $L(\bbTheta)$ in \eqref{e:loss}. Parameter efficiency is a well-documented feature of unrolled architectures~\cite{monga2021spmag}. Further training details, including the specification of the loss and hyperparameter choices, are outlined in the numerical evaluation Section \ref{S:simulation} and in Appendix \ref{app:pre-train}.

\section{Numerical Evaluation}\label{S:simulation}

We perform a comprehensive numerical evaluation to assess the recovery performance and computational efficiency of the proposed SLoG-Net architecture. We test the model in various instances of the source localization task described in Section \ref{ssec:prob_statement}. First we run simulated tests (see Section \ref{Ss:general_exp_setting} for a general description of the experimental setting) and compare SLoG-Net against the iterative ADMM algorithm in Section \ref{Ss:compare_ADMM}, across different types of graphs (Section \ref{Ss:Different_graphs}), and against a selection GNN~\cite{gama2019tsp} in Section \ref{Ss:Compare_CrsGNN}. Finally, in Section \ref{Ss:compare_IVGD} we examine a real data scenario by leveraging the Digg 2009 data set~\cite{hogg2012social}. For this challenging task, we compare SLoG-Net against the Invertible Validity-aware Graph Diffusion (IVGD) NN approach for source localization in~\cite{wang2022invertible}. The Python notebook with the code used to obtain the experimental results reported here is publicly available at \url{https://hajim.rochester.edu/ece/sites/gmateos/code/SLoG-Net.zip}.

\subsection{General experimental settings} \label{Ss:general_exp_setting}

\noindent \textbf{Synthetic data generation.} The graph shift operator is selected as the degree-normalized adjacency matrix $\bbS = \textrm{diag}^{-\frac{1}{2}}(\bbA \mathbf{1}_N) \cdot \bbA \cdot \textrm{diag}^{-\frac{1}{2}}(\bbA \mathbf{1}_N)$. 
With $\ccalT$ denoting the training set, the sparse sources $\bbX\in\mathbb{R}^{N\times |\ccalT|}$ are drawn from the Bernoulli-Gaussian model; e.g.,~\cite{chang2024exact}. Specifically, $\bbX = \mathbf{\Omega} \circ \bbR$, where $\mathbf{\Omega} \in \reals^{N \times |\ccalT|}$ has i.i.d. entries $\Omega_{ij}\sim \textrm{Bernoulli}(\theta)$, with sparsity level $\theta=0.15$; and $\bbR \in \reals^{N \times |\ccalT|}$ is a random matrix (independent of $\mathbf{\Omega}$) with i.i.d. entries $R_{ij}\sim \textrm{Normal}(0,1)$. Unless otherwise stated, realizations of filter coefficients are generated as $\bbh= (\bbe_1 + \varphi \bbb)/\|\bbe_1 + \varphi \bbb\|_1$, where $\bbe_1=[1,0,\ldots,0]^\top \in \mathbb{R}^L$ is the first canonical basis vector and $\bbb\sim \textrm{Normal}(\mathbf{0}_L,\bbI_L)$. The analysis in~\cite{chang2024exact} suggests that recovery is harder for ``less-impulsive'' filters, so we henceforth stick to a challenging instance where $\varphi=1$. The filter order is chosen to be $L=5$. 

For each training epoch, the training samples in $\ccalT$ are randomly split into $Q$ mini-batches of 
$P= |\ccalT|/Q$ signals, namely $\{\bbX_{q}\}_{q=1}^{Q}\in\mathbb{R}^{N\times P}$. We sample $Q$ graph filter coefficients $\{\bbh_q\}_{q=1}^{Q}$ ($L=5$, $\varphi = 1$) and randomly assign them to the input signal mini-batches to generate the observations $\bbY_{q} = \bbV\text{diag}(\bbPsi_L \bbh_q)\bbV^\top\bbX_q$,  $q=1,\ldots,Q$. We use a validation set $\bbX_{\textrm{val}}$ (Bernoulli-Gaussian with $\theta=0.15$) of size $P_{\textrm{val}} = P$, with observations $\bbY_{\textrm{val}} = \bbV\text{diag}(\bbPsi_L \bbh_{\textrm{val}})\bbV^\top\bbX_{\textrm{val}}$, where $\bbh_{\textrm{val}}$ is drawn from the same distribution as $\{\bbh_q\}_{q=1}^{Q}$.\vspace{2pt}

\noindent \textbf{Graphs.} In the following experiments, we implement SLoG-Net and compare it with other approaches using various undirected and unweighted random graphs, as well as real-world networks. In Section \ref{Ss:compare_ADMM}, we use Erd\H{o}s-R\'{e}nyi (ER) random graphs with $N = 20$ nodes and edge formation probability $p=0.3$. In Section \ref{Ss:Different_graphs}, we examine SLoG-Net's recovery performance across various graph ensembles, including: (i) ER ($N = 20$, $p = 0.3$); (ii) stochastic block model (SBM) with $N = 20$ nodes and $N_C = 3$ communities (with edge probabilities $p_\textrm{within}= 0.8$, $p_\textrm{between}=0.2$); (iii) Barabási–Albert (BA) with $N = 20$ nodes; (iv) random geometric (RG) with $N = 20$ nodes and critical distance $r_\textrm{cri}=0.2$; and (v) real-world social networks such as dolphins ($N = 62$) and Zachary's karate club ($N = 34$). In Section \ref{Ss:Compare_CrsGNN}, we evaluate SLoG-Net on the same SBM graph used in Section \ref{Ss:Different_graphs}. In Section \ref{Ss:compare_IVGD}, we use sub-graphs with $N=20$ nodes randomly sampled from the Digg friendship network \cite{hogg2012social}. These sub-graphs and their construction are discussed in further detail in Section \ref{Ss:compare_IVGD}. 
\vspace{2pt}

\noindent \textbf{Training method.} We train SLoG-Net with $K = 5$ layers and use the relative root mean square error (RE) of $\bbX$ as loss function. Notice that if $\{\hbX,\hbh\}$ is a solution to the bilinear problem, then so is $\{-\hbX, -\hbh\}$ and accordingly we minimize
\begin{equation*}
{\small
L(\bbTheta) = \sum\limits_{q=1}^{Q} \min \left( \frac{\| \Phi(\bbY_q;\bbTheta) - \bbX_q \|_F}{\|\bbX_q\|_F},  \frac{\| \Phi(\bbY_q;\bbTheta) + \bbX_q \|_F}{\|\bbX_q\|_F}\right)
}
\end{equation*}
using the Adam optimizer~\cite{kingma2014adam} implemented in PyTorch. 

We initialize $\{\rho_1^{(k)},\rho_2^{(k)},\tau^{(k)}\}_{k=1}^K$ as i.i.d. samples from the uniform distribution in $[0,1]$, since these parameters are constrained to be non-negative. All other parameters in $\bbTheta$ are randomly drawn from a standard Gaussian distribution.
We consider $30$ epochs for training. In each epoch, we estimate the sparse sources $\{\Phi(\bbY_q;\hat{\bbTheta}_q)\}_{q= 1}^{Q}$ using the training batches $\{\bbY_q ; \bbX_q\}_{q= 1}^{Q}$. We choose one batch out of every 20 batches to compute the loss on the validation set $\{\bbY_{\textrm{val}};\bbX_{\textrm{val}}\}$ and record both the value of the loss and the network parameters. In the end, we select the model $\hat{\bbTheta}$ that has minimum validation loss across the entire training process.\vspace{2pt}

\noindent \textbf{Testing protocol and figures of merit.} For testing, we sample an independent test set $\{\bbX_{\textrm{test}},\bbh_{\textrm{test}}\}$, where $\bbX_{\textrm{test}} \in\mathbb{R}^{N\times P_{\textrm{test}}}, P_{\textrm{test}} = P$. We generate diffused signals $\bbY_{\textrm{test}} = \bbV\text{diag}(\bbPsi_L \bbh_{\textrm{test}})\bbV^\top\bbX_{\textrm{test}} + \eta\bbN$, where $\bbN\sim\text{Uniform}(-1,1)^{N\times P_\textrm{test}}$ and $\eta$ is the noise level. We do a forward inference pass through the trained SLoG-Net model to generate predictions $\hbX=\Phi(\bbY_{\textrm{test}};\hat{\bbTheta})$, and use the ground-truth sources $\bbX_{\textrm{test}}$ to assess recovery performance. Naturally, the quality of the estimated inverse filter frequency response $\hat{\tilde{\bbg}}_{\textrm{test}}$ can be evaluated as well. 

For performance assessment, we consider two figures of merit. Firstly, we evaluate the test error $\textrm{RE}=\|\Phi(\bbY_{\textrm{test}};\hat{\bbTheta}) - \bbX_{\textrm{test}}\|_F/\| \bbX_{\textrm{test}}\|_F$. We also compute the accuracy (ACC) in recovering the support of $\bbX_{\textrm{test}}$, i.e., the source locations. To identify the support, we introduce a thresholding approach with threshold $\kappa=10^{-1}$. If a predicted entry satisfies $|[\Phi(\bbY_{\textrm{test}};\hat{\bbTheta})]_{ij}|\geq \kappa$, the index pair $(i,j)$ will be considered a member of the estimated support $\text{supp}_{\kappa}(\cdot)$. Accordingly,  the recovered sources are $\hat{\ccalI}_{\textrm{test}}:=\text{supp}_{\kappa}(\Phi(\bbY_{\textrm{test}};\hat{\bbTheta}))$. We also apply the threshold to the ground-truth sources, so the sought support set is $\ccalI_{\textrm{test}}:=\text{supp}_{\kappa}(\bbX_{\textrm{test}})$.\vspace{2pt}

\begin{figure}[t]
	\centering    
	{\includegraphics[width=\linewidth]{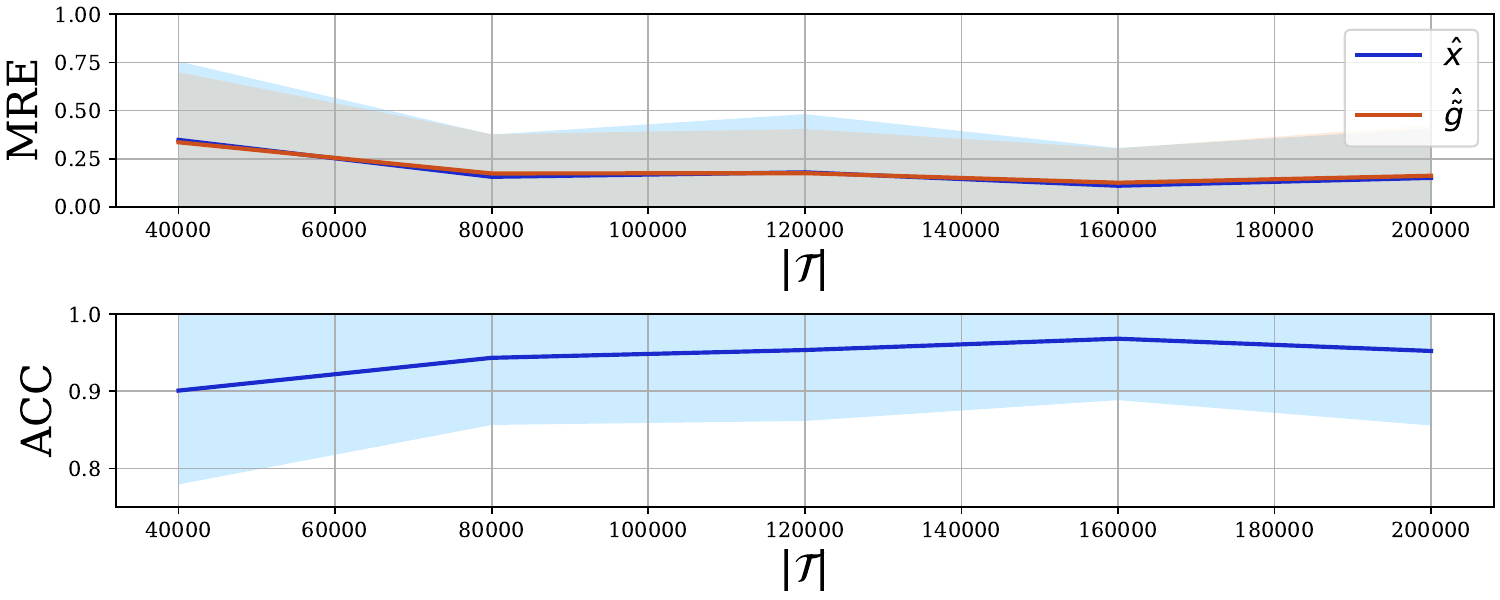}}
	\caption{Recovery performance vs. training set size $|\ccalT|$. (top) Mean test relative error (MRE) of the recovered source signal $\hat{\bbX}$ (blue) and estimated inverse filter frequency response $\hat{\tilde{\bbg}}$ (red), respectively. The shaded region represents the estimated standard error, after averaging over 10 realizations. (bottom) Mean accuracy (ACC) of source support estimation and estimated standard deviation. The best performance is attained for $|\ccalT|\geq 160\textrm{k}$, but gains are marginal beyond $80\textrm{k}$ signals.}
	\label{fig:slog_nTrain}
\end{figure}

\noindent \textbf{Determining the training set size.} To explore the relation between recovery performance and and the size of training set $|\ccalT|$, we train SLoG-Net on a ER graph ($N=20$, $p=0.3$) with different $|\ccalT|$, and compute RE and ACC on the test set. We try $|\ccalT|\in\{40\textrm{k},\ldots,200\textrm{k}\}$, with fixed minibatch size $P = 400$. For each $|\ccalT|$, the experiment is repeated 10 times. As illustrated in Fig. \ref{fig:slog_nTrain} (top), we find that the mean RE initially decreases when $|\ccalT|$ increases, and then it becomes stable when $|\ccalT|\geq 160$k. This is consisent with Fig. \ref{fig:slog_nTrain} (bottom), which shows ACC reaches a maximum value when $|\ccalT|\approx 160\textrm{k}$. At least in this setting, the gains are marginal beyond $|\ccalT|\approx 80\textrm{k}$. In order to make sure that the SLoG-Net is trained well, we henceforth use $|\ccalT| = 200$k as default for the rest experiments.

The minibatch size $P$ also affects the training process. As discussed in~\cite{chang2024exact}, $P$ naturally drives the successful recovery rate
of the convex relaxation \eqref{e:opt_prob_convex}. Generally, a larger network $(N)$ and/or denser sources $(\theta)$ require larger $P$. But when the total number of training samples $|\ccalT|$ is fixed, a larger $P$ might challenge the training process as the number $Q$ of minibatches drops (increasing variability). Our results show that $P=400$ is sufficient for SLoG-Net to train stably on graphs with $N=20$ nodes, yielding satisfactory test performance; see Fig. \ref{fig:slog_nTrain}. 
\vspace{2pt}

\noindent \textbf{Hyperparameter selection.} The SLoG-Net architecture has two hyperparameters: the number of layers $K$ and the number of columns in $\bbM$, i.e., $d$. To balance network complexity and recovery performance, we found that $K=5$ is the optimal number of layers after experimenting with different values through a grid search. For $d$, a larger value implies $\tbg$ is constrained to a smaller subspace with more parameters to learn. Our numerical tests have shown that performance improves markedly when going from $d=1$ to $2$, with diminishing returns for $d\geq 2$ given the increase in complexity. As a result, we use $d=2$ for all subsequent experiments.



\subsection{Comparisons with the ADMM algorithm} \label{Ss:compare_ADMM}

We compare SLoG-Net with the iterative ADMM algorithm in Section \ref{Ss:ADMM}. We examine their recovery performance and inference times during testing, studying the effect of different noise levels $\eta$, number of signals $P$, and number of nodes $N$. 
\vspace{2pt} 

\begin{figure}[t]
	\centering    
	{\includegraphics[width=\linewidth]{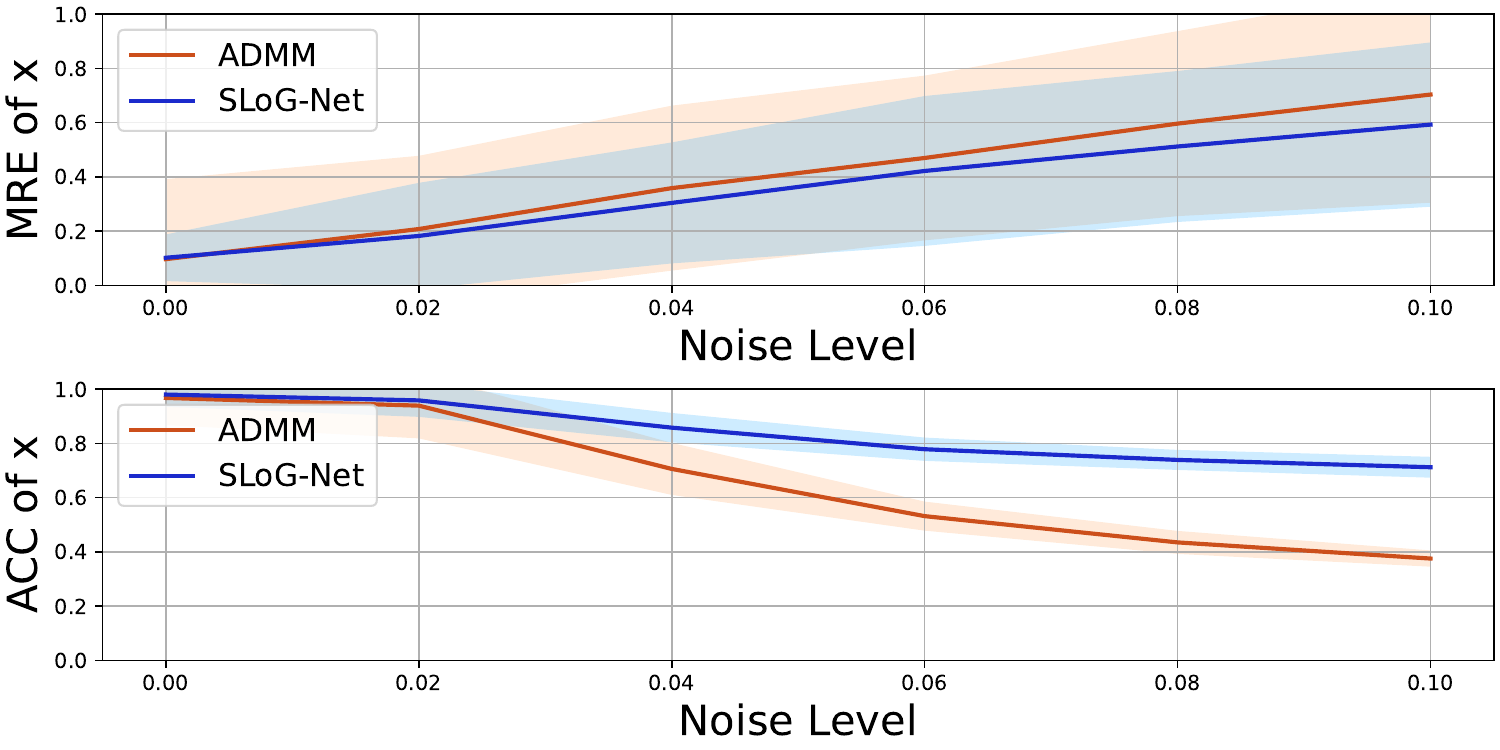}}
	\caption{Recovery performance of SLoG-Net vs. ADMM for different noise levels. (top) Test MRE of the recovered source signal $\hat{\bbX}$ estimated via iterative ADMM (red) and SLoG-Net (blue), as a function of $\eta$. (bottom)) Mean ACC of support estimation for both methods. The shaded region represents the estimated standard error, after averaging over 10 realizations. Performance degrades gracefully for both approaches, but SLoG-Net exhibits better robustness.}
	\label{fig:slog_vs_admm_noise}
\end{figure}

\noindent \textbf{Noise level $\eta$.} To explore the robustness of SLoG-Net in the presence of additive noise corruptin $\bbY$, we first train the model with noise-free data ($\eta=0$) as outlined in Section \ref{Ss:general_exp_setting}. Then we generate test sets with different noise level $\eta$ and evaluate the recovery performance. We also run ADMM (Algorithm 1) on the same test sets and we compare the mean RE and ACC of both approaches (averaged over $10$ independent realizations). Fig. \ref{fig:slog_vs_admm_noise} shows SLoG-Net achieves lower RE than ADMM as the noise level increases. We also find SLoG-Net still attains a high ACC even when $\eta > 0.06$. While the performance of both approaches degrades gracefully (see~\cite{chang2024exact} for noise stability results of the convex relaxation we solve via ADMM), SLoG-Net exhibits higher tolerance to noise in this setting. In addition, the mean wall-clock inference time for SLoG-Net, averaged over 10 realizations, is around 0.009s, uniformly across different noise levels. On the other hand, the mean elapsed time for ADMM, averaged over 10 realizations, is 1.990s at $\eta = 0$ and 7.420s at $\eta = 0.1$. We find ADMM requires more iterations to attain convergence when the noise added to the observations increases. 

\begin{figure}[t]  
	\begin{minipage}[b]{0.24\textwidth}
		\centering
		\includegraphics[width=1\linewidth]{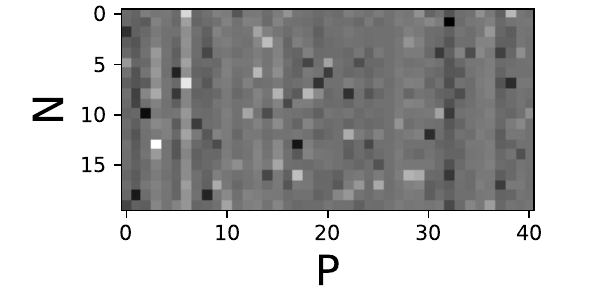}
		\centerline{(a)}\medskip
	\end{minipage}
	\hfill
	\begin{minipage}[b]{0.24\textwidth} 
		\centering
		\includegraphics[width=1\linewidth]{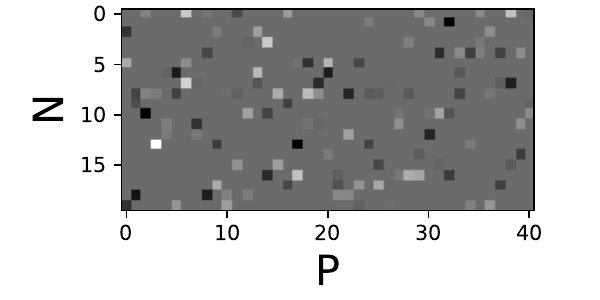}
		\centerline{(b)}\medskip
	\end{minipage}
     \vskip\baselineskip
     \vspace{-10pt}
	\begin{minipage}[b]{0.24\textwidth} 
		\centering
		\includegraphics[width=1\linewidth]{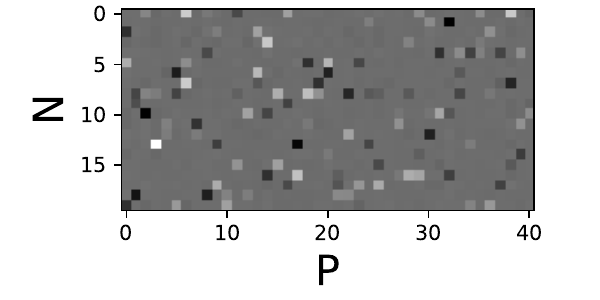}
		\centerline{(c)}\medskip
	\end{minipage}
	\hfill
	\begin{minipage}[b]{0.24\textwidth} 
		\centering
		\includegraphics[width=1\linewidth]{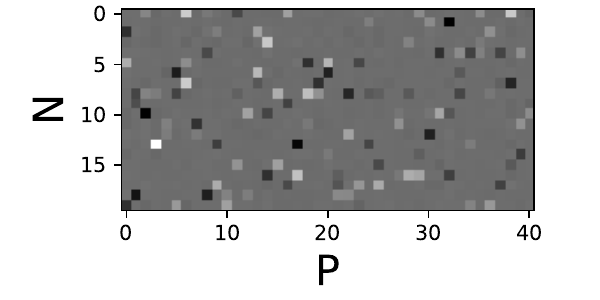}
		\centerline{(d)}\medskip
	\end{minipage}
    
     \vskip\baselineskip
     \vspace{-10pt}
     \centering
	\begin{minipage}[b]{0.36\textwidth} 
		\centering
		\includegraphics[width=1\linewidth]{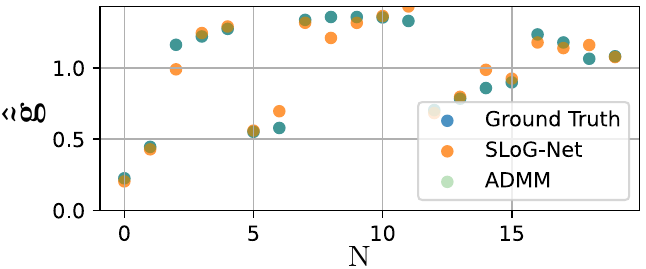}
		\centerline{(e)}\medskip
	\end{minipage}
    \vspace{-10pt}
\caption{A visual comparison of SLoG-Net and ADMM for a representative test realization in the source localization task. (a) Observations $\bbY$; (b) ground-truth sources $\bbX_{\textrm{test}}$; (c) SLoG-Net source estimates; (d) ADMM source estimates. (e) Recovered inverse filter $\hat{\tbg}$: ground truth (blue), SLoG-Net (orange), ADMM (green). For (a)-(d), only $P_1=41$ columns are shown out of a total $P=400$. SLoG-Net's ability to generate accurate predictions (approximating ADMM's model-based solution) is apparent.}\label{fig:visual} 
\end{figure}

For a qualitative assessment, estimation results for a representative test realization when $\eta = 0$ are shown in Fig.~\ref{fig:visual}. In the interest of space, we depict the first $P_1=41$ (out of $P=400$) columns of the observation matrix $\bbY$, ground-truth sources $\bbX_{\textrm{test}}$, as well as SLoG-Net and ADMM source estimates $\hbX$ in Figs.~\ref{fig:visual}(a) - (d), respectively. In Fig.~\ref{fig:visual}(e), the true inverse filter frequency response (blue), and the corresponding estimates obtained via SLoG-Net (orange) and ADMM (green) are presented. We find SLoG-Net recovers the source signals and inverse filter fairly accurately in the absence of noise, offering reasonably good approximations to the ADMM solutions. 
\vspace{2pt}

\noindent \textbf{Number of nodes $N$.} We conducted some timing experiments for ER graphs with increasing number of nodes $N$. To this end, we trained SLoG-Net models for $N \in \{20,40,\ldots,100\}$, with fixed source sparsity level $\theta = 0.15$ and training set size $|\ccalT| = 200$k. For testing, we let $P_\textrm{test} = P = 400$ in all cases. While the recovery error attained by SLoG-Net naturally increases with graph size $N$ (the problem becomes more challenging and we do not add more training data or increasing the test batch size $P_\textrm{test}$), the timing results in Table~\ref{table:inference_time} -- especially when compared to ADMM -- are telling. Indeed, notice how SLoG-Net's mean inference time remains fairly invariant and around $10^{-2}\textrm{s}$. On the other hand, ADMM scales worse with $N$ and is typically a couple of orders of mangnitude slower when it comes to obtaining a solution. Granted, SLoG-Net's extra training time is not accounted for here -- we wish to highlight the computational efficiency of an unrolling during inference.   

\begin{table}
\centering
\begin{tabular}{ |p{0.4cm}||p{2.0 cm}|p{0.9 cm}|  }
 \hline
$N$ & \centering SLoG-Net & ADMM\\
 \hline
 20 & $0.95\times 10^{-2} $    & 2.04 \\
 40 &  $1.09\times 10^{-2}$  & 5.70   \\
 60 &  $1.27\times 10^{-2}$  & 9.41 \\
 80 &  $1.42\times 10^{-2}$  & 12.29  \\
100 &  $1.64\times 10^{-2}$  & 14.62 \\
 \hline
\end{tabular}
\caption{\label{table:inference_time} Mean inference time (sec.). Comparison between SLoG-Net and the ADMM solver for different $N$, with $P = 400$.}
\end{table}

\begin{figure}[t]
	\centering    
	{\includegraphics[width=\linewidth]{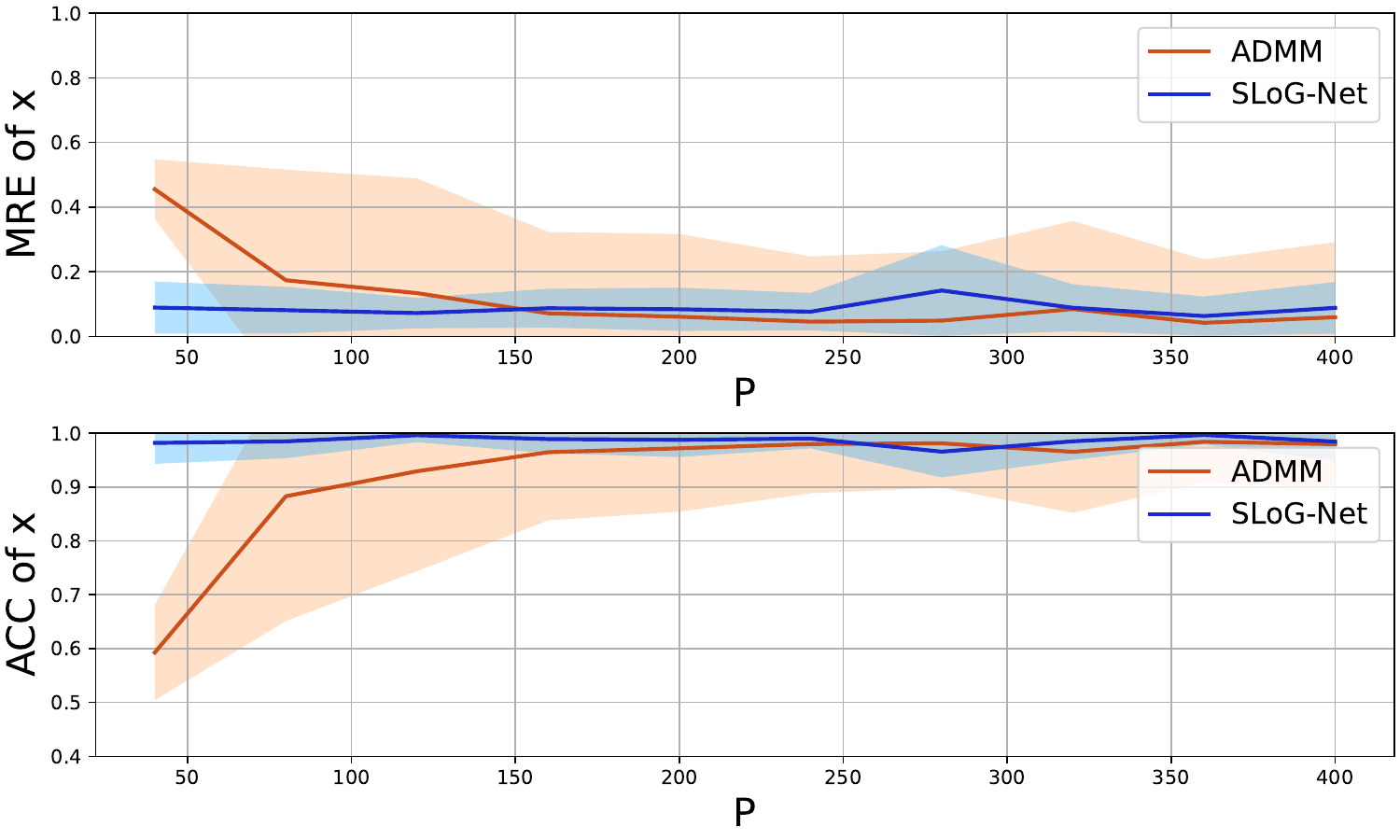}}
	\caption{Recovery performance of SLoG-Net vs. ADMM for different number of signals. (top) Test MRE of the recovered source signal $\hat{\bbX}$ estimated via iterative ADMM (red) and SLoG-Net (blue), as a function of $P$. (bottom) Mean ACC of support support estimation for both methods. The shaded region represents the estimated standard error, after averaging over 10 realizations. SLoG-Net can outperform ADDM, especially when the number of signals is smaller because it benefits from training, and exhibits reduced variability.}
	\label{fig:slog_vs_admm_P}
\end{figure}

\noindent\textbf{Observation size $P$.} Finally, we compare the recovery performance of SloG-Net and  ADMM as a function of the number of observations $P$. The training minibatch size remains equal to $P$ and other experiment parameters are kept fixed, e.g., $N = 20$, $S = \theta N = 3$. We tested $P\in\{ 40, 80,\ldots, 400\}$ and the results are shown in Fig.\ref{fig:slog_vs_admm_P}. For both the mean RE and the ACC, SLoG-Net outperforms ADMM when $P< 160$. When $P\geq 160$, the iterative ADMM achieves similar (or marginally better) mean RE and ACC than SLoG-Net, but the latter exhibits reduced variability across realizations. In terms of timing, the mean inference time for SLoG-Net is around 0.009s across the range of $P$ values; while for ADMM it is 8.412s at $P = 40$, it decreases to 0.838s at $P = 160$, and then it increases to 1.851s at $P = 400$.  When $P$ is too small, it is harder to obtain a good solution via the relaxation \eqref{e:opt_prob_convex}, and ADMM may struggle to converge. However, SLoG-Net has stable recovery performance for different $P$ because it benefits from an additional training phase. 

\subsection{Recovery performance on different graphs} \label{Ss:Different_graphs}

We also study the efficacy of SLoG-Net in identifying the sources across various graph ensembles, including random graphs ($N = 20$) such as ER, SBM, RG and BA, as well as the real karate club graph ($N=34$), and the dolphins social network ($N=62$). We use the settings and methods described in Section \ref{Ss:general_exp_setting}. In Table \ref{table:slog_diff_graphs} we report the mean RE of $\hat\bbX$ and $\hat\tbg$, as well as the support recovery ACC. Despite the relatively high RE ($>0.3$) for the source signal $\hat\bbX$ or the inverse filter $\hat\tbg$ for some of the graphs (especially the more structured and larger ones), the support estimate ACC remains high ($>0.8$). The result has a twofold interpretation. First, SLoG-Net's subpar RE performance on certain graphs may be due to $\tbg$ having a significant component that is orthogonal to $\textrm{span}(\mathbf{1}_N)$. Indeed, results in~\cite{chang2024exact} show that $\|\text{P}_1^\perp\tbg \|_2$ (where $\text{P}_1^\perp := \bbI_N-\frac{1}{N}\mathbf{1}\mathbf{1}^\top$ is the projector onto $\textrm{span}^\perp(\mathbf{1}_N)$) can be viewed as a condition number of the problem \eqref{e:opt_prob_convex}). Hence, the higher variability (the eigenvalue distribution of different graph types affects the Vandermonde matrix $\bbPsi_L$) of randomly generated filters $\tbg$ for some graphs, may increase the problem difficulty that manifests through higher REs. But SLoG-Net is designed to optimize a more flexible version of \eqref{e:opt_prob_convex}, with a learnable constraint. Using $\bbM$ instead of $\mathbf{1}_N$ affects the problem's conditioning, (we conjecture) likely contributing to keep ACC rates at satisfactory levels. A more in-depth analysis is certainly of interest, but beyond the scope of this paper.
\begin{table}
\centering
\begin{tabular}{ |p{1.5cm}||p{0.4cm}|p{1cm}|p{1.3cm}|p{1.2cm}|p{0.7cm}|  }
 \hline
Graph & $N$ & $\|\bbP^\perp_1 \tbg\|_2$ &MRE of $\hat{\bbX}$ & MRE of $\hat{\tilde{\bbg}}$&ACC\\
 \hline
 \hline
ER & 20 & 6.885    & 0.149    & 0.164 &   0.953 \\
SBM & 20 & 7.806  & 0.219  & 0.215  & 0.914\\
RG & 20 & 7.591 & 0.383 & 0.377 &  0.869\\
BA & 20 & 14.547 & 0.579 & 0.537 & 0.772\\
karate & 34 & 23.996 & 0.454 & 0.452 & 0.958\\
dolphins & 62 & 39.254  & 0.719  & 0.578 & 0.841\\
 \hline
\end{tabular}
\caption{\label{table:slog_diff_graphs}SLoG-Net recovery performance for different graphs.}
\end{table}

\subsection{Comparison with a GNN approach} \label{Ss:Compare_CrsGNN}

Here we explore the feasibility of using SLoG-Net for community detection in an SBM with $N_c$ communities. Our goal is not to demonstrate state-of-the-art performance in this well-investigated task, but rather to find an application domain where comparison with the GNN models in~\cite{gama2019tsp} is feasible. So far, the support of each $\bbx_i$ was specified via i.i.d Bernoulli random variables, in community detection all of the sources (the $S$ non-zero entries of $\bbx_i$) are drawn randomly within a single subset of indices from $\{1,\ldots,N\}$, representing the members of the community to be identified. So we cast this simple version of community detection as \emph{structured} source localization, where active sources can be located in only one of the $N_c$ communities. 

To illustrate this further, we consider an SBM graph with $N = 20$ nodes and $N_c = 3$ communities. Each of the input signals $\bbx_i$ are generated as follows: (i) select a community by drawing an integer $c_i$ uniformly at random from $\{1,\ldots,N_c\}$; (ii) randomly select $S = \theta N = 3$ nodes among the members of the selected community $c_i$ -- the active sources which we encode in the support vector $\bbomega_i\in\{0,1\}^N$; (ii) compute $\bbx_i = \bbr_i \circ \bbomega_i$, where $\bbr_i\sim \textrm{Normal}(\mathbf{0}_N,\bbI_N)$. All in all, $\bbx_i$ are samples from a variation of a Bernoulli-Gaussian model, whose support is constrained to a randomly-chosen subset of nodes (the members of the chosen community $c_i$).\vspace{2pt}




\noindent \textbf{Experimental details.}
For training, we generate $|\ccalT| = 200$k input signals $\bbX\in\mathbf{R}^{N\times |\ccalT|}$ and source community labels $\bbK\in\{0,1\}^{N_c\times |\ccalT|}$, via one-hot encoding of each $c_i$. We train SLoG-Net as described in \ref{Ss:general_exp_setting}. For the baseline selection GNN model~\cite{gama2019tsp}, we generate $|\ccalT|$ graph filters $\{\bbh_i\}_{i = 1}^{|\ccalT|}$ and assign them one to one to the input signal $\{\bbx_i\}_{i = 1}^{|\ccalT|}$ to generate the observations $\{\bby_i\}_{i = 1}^{|\ccalT|}$, mimicking the setting in~\cite[Sec. V-A]{gama2019tsp}. The GNN is trained with supervised data $\{c_i,\bby_i\}_{i = 1}^{|\ccalT|}$. For testing, we generate one test set of $P_{\textrm{test}} =P= 400$ samples, i.e., $\bbX_\textrm{test}\in\mathbf{R}^{N\times P}$, a graph filer $\bbh_\textrm{test}$, and observations $\bbY_\textrm{test} = \bbV \text{diag}(\bbPsi_L \bbh_\textrm{test})\bbV^\top \bbX_\textrm{test}+\eta\bbN$, for different noise levels $\eta \in\{0,0.02,\ldots,0.1\}$. Notice that our goal is to estimate the source community as in~\cite[Sec. V-A]{gama2019tsp}, not the source nodes. But since SLoG-Net recovers the input signal $\hbX$, we apply a strategy that considers the community, where the entry with highest magnitude of the estimated source $\hat{\bbx}_i$ resides, as SLoG-Net's community estimate. \vspace{2pt}

\begin{figure}[t]
	\centering    
	{\includegraphics[width=\linewidth]{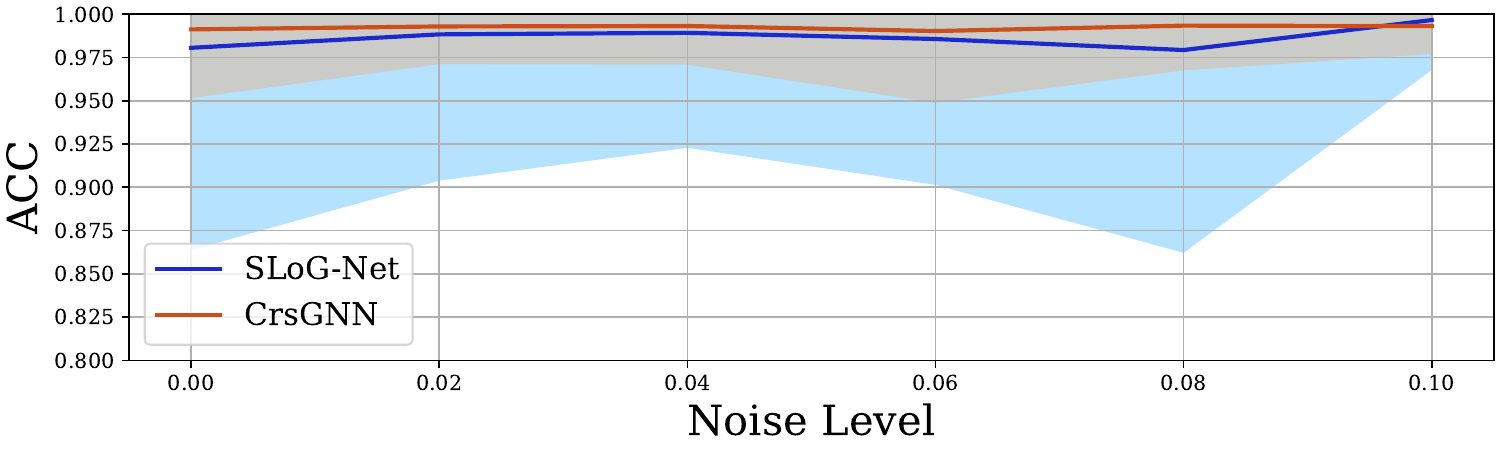}}
	\caption{Recovery performance of SLoG-Net vs. GNN for different noise levels. Mean ACC of estimated communities estimated via GNN~\cite{gama2019tsp} (red) and SLoG-Net (blue) as a function of $\eta$. The shaded region represents the estimated standard deviation, after averaging over 10 realizations. SLoG-Net is trained on a harder (higher-resolution) source localization task, without community label supervision -- which the GNN model in~\cite{gama2019tsp} struggles to solve. Still, SLoG-Net attains competitive community detection ACC rates.}
	\label{fig:slog_vs_crsgnn_noise}
\end{figure}

\noindent \textbf{Results and discussion.} We compare the mean community detection ACC for both SLoG-Net and the GNN; the results are reported in Fig. \ref{fig:slog_vs_crsgnn_noise}. Apparently, the GNN attains very high ACC rates for the entire noise level spectrum. Remarkably, SLoG-Net achieves competitive performance that is robust across noise levels. Notice that SLoG-Net is trained on a harder source localization task (node identification) and without community label supervision, which the GNN model in~\cite{gama2019tsp} struggles to solve. To carry out the comparison, we settled on community detection, a lower resolution source localization variant that favors the GNN approach.

\subsection{Real-data experiment} \label{Ss:compare_IVGD}

In this last section, we test SLoG-Net on a source localization problem we set up using the Digg 2009 data set~\cite{hogg2012social}. \vspace{2pt}

\noindent \textbf{Data preprocessing.} The Digg 2009 data set consists of vote records and a social network of users.  The vote records contain $3$M votes from $139$k users on $3553$ popular stories, along with the voting timestamps. The social network of users includes $1.7$M friendship links between $71$k unique users. 
To assess SLoG-Net's source localization capabilities in a  real-world setting, we treat the voting history of each story as a signal diffused over the social network graph $G$. However, processing a graph with 139k or 71k users is infeasible for SLoG-Net as there are only at most $3553$ training and testing samples. To ensure that the graph is connected and not too large so that it can be processed for effective learning, 
we randomly selected $N=20$ users with the following criteria: (i) they must have cast at least 100 votes; (ii) all of their friendship links are mutual; and (iii) their friendship subgraph is connected. Following these guidelines, we randomly sampled 5 subgraphs with $N=20$ nodes from Digg, with mean degree $16.0\pm 0.74$ and $P_\textrm{story}=3348.0\pm 55.7$ stories voted by each user.

To generate the graph signals, we considered each story as a sample, using the 10\% earliest votes as the binary sources $\bbx_p$ and all votes as the observations $\bby_p$. Because the sources $\bbX\in\{0,1\}^{N\times P_\textrm{story}}$ are a subset of the observations $\bbY\in\{0,1\}^{N\times P_\textrm{story}}$, we only consider identifying the sources from the support set of the observations $\textrm{supp}(\bbY)$.\vspace{2pt}

\noindent \textbf{Experimental details.} The challenge of localizing sources using SLoG-Net on the sampled Digg data is twofold. First, the sample size is limited, i.e., $|\ccalT|\approx 3.5$k, challenging effective training; secondly, both the sources $\bbX$ and observations $\bbY$ are binary. However, if we focus on recovering 
$\textrm{supp}(\bbX)$, SLoG-Net may still yield useful results given the binary observations $\bbY$. To address these challenges, we adopted several strategies. 

First, we found that selecting a proper mini-batch size $P$, along with an appropriate observation size $P_\textrm{test}$, was crucial for both training and testing. After several attempts, we determined that while a larger training batch size $P$ would improve the performance of SLoG-Net, it would also result in a smaller training set size $|\ccalT|$, as we wanted to use $P_\textrm{test}=P$ for consistency. Therefore, we opted for $P=400$, which set $P_\textrm{test} = 400$ and $|\ccalT| = P_\textrm{story} - P_\textrm{test}$, so $|\ccalT|\approx 2.9$k on average.

\begin{figure}[t]
	\centering    
	{\includegraphics[width=0.8\linewidth]{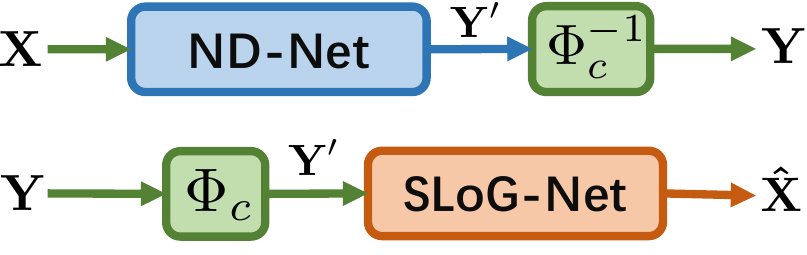}}
	\caption{(top) Pre-training architecture and (bottom) calibrated SLoG-Net.}
	\label{fig:cSLoG-Net}
\end{figure}

To address the second challenge of binary inputs, 
we aim to find some calibration operator $\Phi_c: \bbY \mapsto \bbY'$ that maps the binary observations $\bbY$ to real-valued graph signals $\bbY'$, which can be viewed as the result of diffusing the sparse binary sources $\bbX$. Then we can apply SLoG-Net to predict $\hat\bbX = \Phi(\bbY';\bbTheta)$, as depicted in Fig.~\ref{fig:cSLoG-Net} (bottom). We were inspired by \cite{wang2022invertible}, where an invertible graph diffusion network (IVGD) is proposed based on the invertible residual network (i-ResNet)~\cite{behrmann2019invertible}. We adopt an i-ResNet structure to construct the \emph{invertible} mapping $\Phi_c$, which can be approximated via fix-point iterations from its inverse $\Phi_c^{-1}:\bbY'\mapsto\bbY$; see Algorithm 2. Specifically, we consider $\Phi_c^{-1}(\bbY';\bbTheta_c) = \frac{1}{2}(\bbY'+\textrm{MLP}(\bbY'))$, where $\textrm{MLP}(\cdot;\bbTheta_c)$ is a three-layer, 1000-hidden unit multi-layer perceptron (MLP) with learnable parameters $\bbTheta_c$. We train $\Phi_c^{-1}$ along with a network diffusion NN (ND-Net), as shown in Fig.~\ref{fig:cSLoG-Net} (top). Details of this pre-training process are presented in Appendix~\ref{app:pre-train}.

SLoG-Net is trained using $\ccalT = \{\bbX,\Phi_c^{-1}(\bbY)\}$; see Fig.~\ref{fig:cSLoG-Net} (bottom). Because the complement $\ccalI_Y^c$ of $\ccalI_Y:=\textrm{supp}(\bbY)$ cannot include the sources, we want to penalize $\ccalI_Y$ more. To this end, we use the weighted loss function 
\begin{equation*}
L_{\phi}(\bbTheta) = \sum\limits_{q=1}^{Q} \min \left(L_{q,\phi}^+(\bbTheta),  L_{q,\phi}^-(\bbTheta)\right),
\end{equation*}
where $L_{q,\phi}^{\pm}(\bbTheta):=  \frac{\|[ \Phi(\bbY_q;\bbTheta) \pm \bbX_q ]_{\ccalI_Y}\|_F + \phi \| [\Phi(\bbY_q;\bbTheta) \pm \bbX_q ]_{\ccalI_Y^c}\|_F}{\|\bbX_q\|_F}$ and $\phi = 0.01$. With these adjustments, training proceeds as described in Section \ref{Ss:general_exp_setting}. 

\setcounter{algocf}{1}
\begin{algorithm}[t]\label{alg:comps_net_inv}
	\SetAlgoLined
	\textbf{Require} $\Phi_c^{-1}$ , $\bbY$,  $K_{\max}$ (number of iterations).\\
	\textbf{Initialize} $\bbY'[0] = \bbY$. \\
	\For{$k=1,2,\ldots,K_{\max}$}{
		$\bbY'[k] \gets 2\bbY - \Phi_c^{-1}(\bbY'[k-1]).$
	}
	\textbf{Output} $\bbY'[K_{\max}]=\Phi_c(\bbY)$.
	\caption{Calibration Mapping $\Phi_c$ via Inversion}
\end{algorithm}

For the IVGD baseline, we run the algorithm provided in~\cite{wang2022invertible}. The number of hidden units is chosen to be $50$, consistent with the experimental setting described in \cite{wang2022invertible}.\vspace{2pt}

\begin{figure}[t]
	\centering    
	{\includegraphics[width=\linewidth]{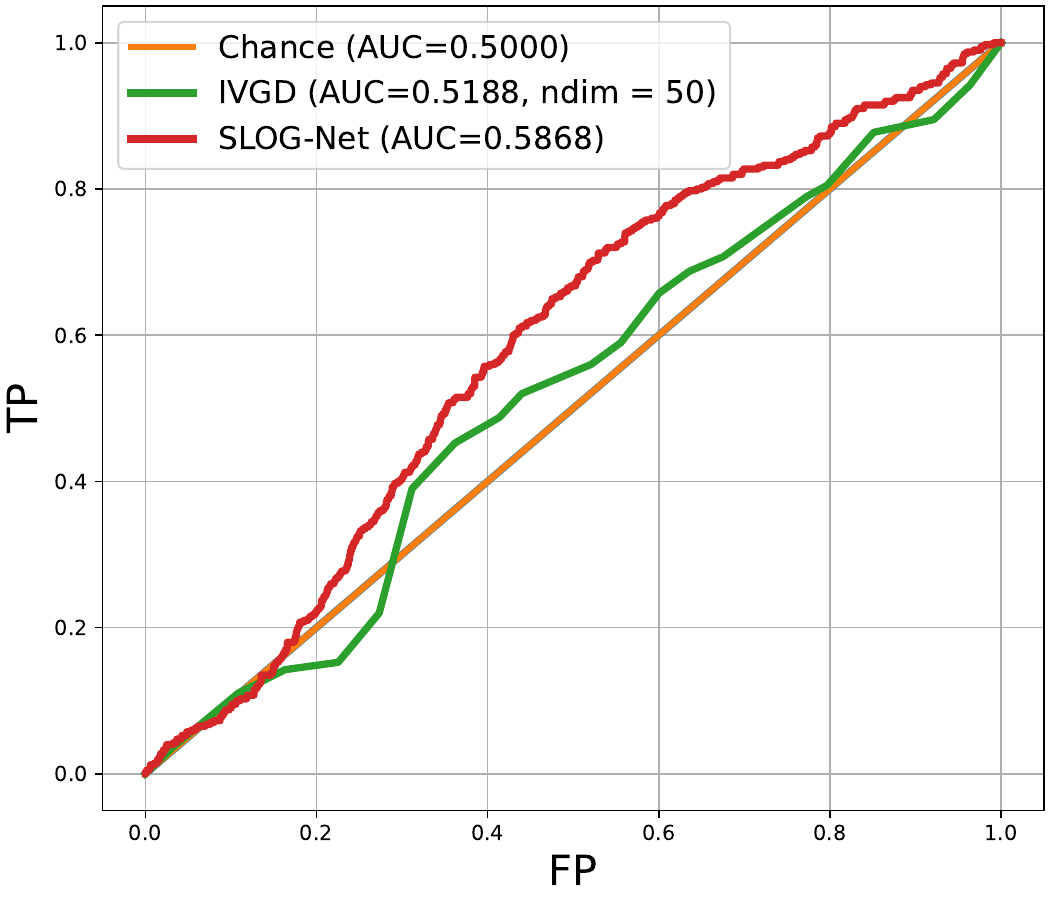}}
	\caption{The ROC curve of SLoG-Net vs. IVGD for one representative sampled subgraph. The mean AUC over 10 different training/testing set realizations are 0.56 and 0.51 for SLoG-Net and IVGD, respectively.}
	\label{fig:slog_vs_ivgd_roc}
\end{figure}

\noindent \textbf{Results and discussion.} To evaluate the source localization performance, we compute the ROC curve and AUC of $\{\hat{\bbX}_{\ccalI_Y}, [\bbX_{\textrm{test}}]_{\ccalI_Y}\}$,  where $\hat{\bbX}$ are the predicted sources, $\bbX_\textrm{test}$ is the ground-truth of the test set and $\ccalI_Y=\textrm{supp}(\bbY_\textrm{test})$. The experiment is repeated twice for each of the $5$ sampled subgraphs, each time the training and test sets are randomly split from all the samples $\{\bbX,\bbY\}$. An ROC generated for one representative sampled subgraph is depicted in Fig. \ref{fig:slog_vs_ivgd_roc}. The mean AUC averaged over 10 realizations is 0.56 and 0.51 for SLoG-Net and IVGD, respectively. While none of the methods perform admirably in this hard problem, SLoG-Net is better at learning representations that are predictive of the sources.

\section{Conclusions and Future Work}\label{S:conclusion}

We developed SLoG-Net, a novel DL approach for source localization on graphs with broader impacts to blind deconvolution of graph signals. The unrolled architecture fruitfully leverages inductive biases stemming from model-based ADMM iterations we also developed, is parameter efficient, fairly robust to noise, and offers controllable complexity after training. Admittedly, there is still work to be done to arrive at a truly scalable solution that is compatible with large-scale problems. Our experimental results with simulated and real network data demonstrate that SLoG-Net exhibits performance on par with the iterative ADMM baseline it is trained to approximate, while attaining order-of-magnitude speedups to generate source location predictions during inference. 

Importantly, SLoG-Net opens the door for further architectural refinements by leveraging advances in optimization, DL, and machine learning on graphs, which we intend to pursue as future work. Exciting ideas include designing a model that fully operates in the vertex domain as well as expanding our performance evaluation protocol to study generalization and transfer to larger graphs, possibly establishing stability and transferability properties of the resulting unrolled (G)NNs. We also look forward to exploring additional application domains in network neuroscience, seismology, and epidemiology. 

\appendix

\subsection{Derivation of the ADMM updates}\label{app:admm}

The ADMM algorithm can be viewed as blending block coordinate-descent (BCD) updates for the primal variables $\{\tbg[k],\bbx[k]\}$, with dual gradient-ascent iterations for the Lagrange multipliers $\{\bblambda[k],\bbmu[k]\}$; see e.g.,~\cite{Bertsekas_Book_Distr,boyd2011distributed,giannakis2016chapter}. Accordingly, when used to solve problem \eqref{e:opt_prob_convex_admm} it entails the following three steps per iteration $k=0,1,\ldots$:

\begin{description}
\item [{\bf [S1]}]  \textbf{Filter updates:}
    \begin{equation}\tilde{\bbg}[k+1] =\argmin_{\tbg}\ccalL_\rho(\bbx[k],\tilde{\bbg},\bblambda[k],\mu[k]).
    \label{S1}\end{equation}

\item [{\bf [S2]}] \textbf{Sources' updates:}
    \begin{equation}\label{S2}\bbx[k+1]=
   \argmin_{\bbx}\ccalL_\rho(\bbx,\tilde{\bbg}[k+1],\bblambda[k],\mu[k]).\\
    \end{equation}
\item [{\bf [S3]}] \textbf{Lagrange multiplier
    updates:}\begin{align}
    \hspace{-0.8cm}\bblambda[k+1] ={} &  \bblambda[k] + \rho_\lambda(\bbZ\tilde{\bbg}[k+1] - \bbx[k+1]),\label{S3_a}\\
    \hspace{-0.8cm}\mu[k+1]={} & \mu[k] + \rho_\mu(\mathbf{1}_N^\top\tilde{\bbg}[k+1] - c).\label{S3_b}
\end{align}
\end{description}
The Lagrange multiplier updates in [S3] coincide with \eqref{eq:admm_update rule_lambda}-\eqref{eq:admm_update rule_mu}. These correspond to gradient-ascent iterations, since the gradients of the dual function are equal to the respective constraint violations in \eqref{e:opt_prob_convex_admm}~\cite{boyd2011distributed}.

What remains is to show that [S1]-[S2] can be simplified to
\eqref{eq:admm_update rule_g}-\eqref{eq:admm_update rule_x}. Starting with [S1], note that the augmented Lagrangian is a strictly-convex, smooth quadratic function with respect to $\tbg$. The gradient of \eqref{e:opt_prob_convex_lagrangian} is
\begin{align*}
\nabla_{\tbg}\ccalL_\rho(\bbx,\tilde{\bbg},\bblambda,\mu) = {}& \rho_{\lambda} \bbZ^\top(\bbZ\tilde{\bbg} -\bbx + \bblambda/\rho_{\lambda} )\\
{}&+ \rho_{\mu}\mathbf{1}_{N}(\mathbf{1}_{N}^\top\tilde{\bbg} - c + \mu/\rho_{\mu}).
\end{align*}
The unique minimizer of \eqref{S1} satisfies the first-order optimality condition $\nabla_{\tbg}\ccalL_\rho(\bbx[k],\tilde{\bbg},\bblambda[k],\mu[k]) =\mathbf{0}_N$. Solving the linear system of equations immediately yields \eqref{eq:admm_update rule_g}, where $\bbGamma:=\rho_{\lambda} \bbZ^\top\bbZ+ \rho_{\mu}\mathbf{1}_{N}\mathbf{1}_{N}^\top$.

Shifting our focus to [S2], one readily recognizes
\eqref{S2} as the proximal operator of the function $\rho_\lambda^{-1}\|\bbx\|_1$ evaluated at
$\bbZ\tilde{\bbg}[k+1]+\bblambda[k]/\rho_\lambda$. Said proximal operator is a soft-thresholding operator; e.g.~\cite{mardani2012admm}, and the update rule \eqref{eq:admm_update rule_x} follows.

To derive the iterations in Fig. \ref{fig:blockdiag}, which tabulates the ADMM algorithm for the modified formulation with the general constraint $\bbM^\top\tilde{\bbg}=\bbm$, one simply mimics [S1]-[S3] to instead minimize the updated augmented Lagrangian
\begin{align*}
\ccalL_\rho(\bbx,\tilde{\bbg},\bblambda,\bbmu) = {}& 
\|\bbx \|_1 + \frac{\rho_{\lambda}}{2}\| \bbZ\tilde{\bbg} -\bbx + \bblambda/\rho_{\lambda} \|_2^2\\ 
	{}&+ \frac{\rho_{\mu}}{2}\| \bbM^\top\tilde{\bbg} - \bbm + \bbmu/\rho_{\mu}\|_2^2.
\end{align*}
%
\subsection{Diagonal structure of $\bbZ^\top \bbZ$}\label{app:diag_structure}
 
Recall $\bbZ:= \bbY^\top\bbV\odot \bbV \in\mathbb{R}^{NP\times N}$, where $\bbY\in\mathbb{R}^{N\times P}$ and $\bbV = [\bbv_1, \ldots, \bbv_N]\in\mathbb{R}^{N\times N}$. Letting $\tbY=\bbV^\top\bbY\in \reals^{N\times P}$, we have 
\begin{equation*}
    \bbZ =  \tilde{\bbY}^\top \odot \bbV 
    =  \begin{bmatrix} [\tilde{Y}^\top]_{11} \bbv_1\:\: [\tilde{Y}^\top]_{12} \bbv_2 \:\: \ldots \:\: [\tilde{Y}^\top]_{1N} \bbv_N\\ 
    [\tilde{Y}^\top]_{21} \bbv_1\:\: [\tilde{Y}^\top]_{22} \bbv_2 \:\: \ldots \:\: [\tilde{Y}^\top]_{2N} \bbv_N\\ 
    \vdots\\
    [\tilde{Y}^\top]_{P1} \bbv_1\:\: [\tilde{Y}^\top]_{P2} \bbv_2 \:\: \ldots \:\: [\tilde{Y}^\top]_{PN} \bbv_N\\ 
    \end{bmatrix} 
\end{equation*}
from where it follows that
\begin{align}\label{eq:diag_entries}
    [\bbZ^\top \bbZ]_{ij} = {}& \begin{bmatrix} [\tilde{Y}^\top]_{1i} \bbv_i\\
    [\tilde{Y}^\top]_{2i} \bbv_i\\
    \vdots\\
    [\tilde{Y}^\top]_{Pi} \bbv_i
    \end{bmatrix}^\top \cdot \begin{bmatrix} [\tilde{Y}^\top]_{1j} \bbv_j\\
    [\tilde{Y}^\top]_{2j} \bbv_j\\
    \vdots\\
    [\tilde{Y}^\top]_{Pj} \bbv_j
    \end{bmatrix}\nonumber\\
    = {}& \sum_{p=1}^P [\tilde{Y}^\top]_{pi}[\tilde{Y}^\top]_{pj}\bbv_i^\top\bbv_j
    = [\tilde{\bbY} \tilde{\bbY}^\top]_{ij} \delta_{ij},
\end{align}
where $\delta_{ij}=\ind{i=j}$ is the Kronecker delta. Notice that \eqref{eq:diag_entries} follows since the graph-shift operator eigenvectors are orthogonal. All in all, we have shown that $\bbZ^\top\bbZ$ is an $N\times N$ diagonal matrix with diagonal elements $\{\|\bbv_i^\top\bbY\|_2^2 \}_{i=1}^N $. We can thus write $\bbZ^\top\bbZ = \text{diag}(\|\bbv_1^\top\bbY\|_2^2,\ldots,\|\bbv_N^\top\bbY\|_2^2)$.

\subsection{Inverting $\bbZ^\top\bbZ + \rho \mathbf{1}_N\mathbf{1}_N^\top$ via the matrix inversion lemma}\label{app:mil_rank_one}

The Sherman–Morrison–Woodbury formula states
\begin{equation} \label{eq:woodbury_lemma}
    (\bbU + \bbB\bbC\bbD)^{-1} = \bbU^{-1} - \bbU^{-1}\bbB(\bbC^{-1}+\bbD\bbU^{-1}\bbB)^{-1}\bbD \bbU^{-1}.
\end{equation}
To apply this identity to invert $\bbGamma \propto \bbZ\bbZ^\top+ \rho \mathbf{1}_N\mathbf{1}_N^\top$, define $\bbz:= [\|\bbv_1^\top\bbY\|_2^2,\ldots,\|\bbv_N^\top\bbY\|_2^2]^\top\in\reals^N$, and then let $\bbU := \bbZ^\top \bbZ = \textrm{diag}(\bbz)$. Comparing $\bbZ^\top\bbZ + \rho \mathbf{1}_N\mathbf{1}_N^\top$ and \eqref{eq:woodbury_lemma}, we let $\bbB := \mathbf{1}_N$, $\bbD = \mathbf{1}_N^\top$, and $\bbC = \rho$. The matrix sum that is to be inverted in the right-hand-side of \eqref{eq:woodbury_lemma} is a scalar, namely $\bbC^{-1} + \bbD \bbU^{-1}\bbB = \rho^{-1}+ \mathbf{1}_N^\top(\bbz^{-1} \circ \mathbf{1}_N):=\zeta$, where with an abuse of notation we let $\bbz^{-1}:=[\|\bbv_1^\top\bbY\|_2^{-2},\ldots,1/\|\bbv_N^\top\bbY\|_2^{-2}]^\top\in\reals^N$ be the (entrywise) vector reciprocal of $\bbz$. Applying \eqref{eq:woodbury_lemma}, we obtain
\begin{align}
    \left(\bbZ^\top \bbZ + \rho \mathbf{1}_N\mathbf{1}_N^\top\right)^{-1} & = (\bbZ^\top \bbZ)^{-1} - \frac{(\bbZ^\top \bbZ)^{-1}\mathbf{1}_N\mathbf{1}_N^\top(\bbZ^\top \bbZ)^{-1}}{\zeta}\nonumber \\
    {}& = \text{diag}(\bbz^{-1}) - \frac{\text{diag}(\bbz^{-1})\mathbf{1}_N\mathbf{1}_N^\top\text{diag}(\bbz^{-1})}{\zeta}\nonumber \\
    {}& = \text{diag}(\bbz^{-1}) - \frac{ \bbz^{-1} (\bbz^{-1})^\top}{\zeta}. \label{eq:matrix_lemma_inverse_last}
\end{align}
While \eqref{eq:matrix_lemma_inverse_last} is particularly simple when the correction to $\bbZ^\top\bbZ$ is a scaled version of an all-ones matrix, a general rank-one correction of the form $\rho \bbm\bbm^{\top}$ is almost identical. Indeed, one just needs to re-evaluate $\zeta=\rho^{-1}+ \bbm^\top(\bbz^{-1} \circ \bbm)$ and the right-most summand in the third line of \eqref{eq:matrix_lemma_inverse_last} becomes $\zeta^{-1}(\bbm\circ \bbz^{-1})(\bbm\circ\bbz^{-1})^\top$.

The computational complexity of \eqref{eq:matrix_lemma_inverse_last} includes: i) computing $\bbz^{-1}$, the entrywise reciprocal of $\bbz$, $O(N)$ assuming $\bbz$ is given; ii) computing $\zeta = \rho^{-1}+ \bbm^\top(\bbz^{-1} \circ \bbm)$, or the sum $\bbm^\top(\bbz^{-1} \circ \bbm) = \sum_i m_i^2/z_i$, $O(N)$;  iii) computing the outer product $(\bbm \circ \bbz^{-1}) (\bbm\circ\bbz^{-1})^\top$, $O(N^2)$; iv) normalizing by $\zeta$, $O(1)$; and updating the diagonal entries by adding $\text{diag}(\bbz^{-1})$, an extra $O(N)$. 
As a result, the overall computational complexity of inverting $\bbZ^\top\bbZ + \rho \bbm\bbm^\top$ is $O(N^2)$. 

There are no order-wise savings when $\bbm=\mathbf{1}_N$, which is what we require to invert $\bbGamma:=\rho_\lambda\bbZ^\top\bbZ+\rho_{\mu}\mathbf{1}_N\mathbf{1}_N^\top$ in the ADMM update \eqref{eq:admm_update rule_g}, or, $\bbGamma^{(k)} = \bbZ^\top\bbZ+\rho_2^{(k)}\mathbf{1}_N\mathbf{1}_N^\top$ in SLoG-Net's filter sub-layer when the constraint parameters $\bbM^{(k)}$ and $\bbm^{(k)}$ are not learnt~\cite{chang2022eusipco}. When $d=1$, the formula \eqref{eq:matrix_lemma_inverse_last} can also be applied to the refined filter sub-layer \eqref{eq:admm_update rule_g}; see Appendix \ref{app:mil_general} for the general case. To appreciate the overall savings, recall that the computational complexity of inverting a general $N\times N$ matrix is $O(N^\omega)$, where $\omega \in\{2.376, 2.807, 3\}$ for three different kind of algorithms; namely, the Coppersmith–Winograd algorithm, the Strassen algorithm, and Gauss–Jordan elimination, respectively. 

\subsection{Inverting $\bbZ^\top\bbZ + \rho \bbM\bbM^\top$}\label{app:mil_general}

Let $\bbU = \bbZ^\top \bbZ = \textrm{diag}(\bbz)$, $\bbC = \rho\bbI_{d}$, and $\bbB = \bbD^\top = \bbM\in\reals^{N\times d}$. From the matrix inversion lemma \eqref{eq:woodbury_lemma} we have,
\begin{align*}
   (\bbZ^\top \bbZ + \rho \bbM\bbM^\top)^{-1} 
     = {} & \text{diag}(\bbz^{-1})\\
     {}&- \text{diag}(\bbz^{-1})\bbM\bar\bbM^{-1}\bbM^\top\text{diag}(\bbz^{-1}),
\end{align*}
%
%
where $\bar\bbM = \rho^{-1}\bbI_{d} + \bbM^\top\text{diag}(\bbz^{-1})\bbM\in\reals^{d \times d}$. 

\subsection{Pre-training process}\label{app:pre-train}
To learn the inverse calibration mapping $\Phi_c^{-1}: \bbY'\mapsto \bbY$ directly, we we would need a training set $\{\bbY,\bbY'\}$. We model the latent $\bbY'$ as the output of network diffusion process driven by sources $\bbX$, with some unknown graph filter $\bbH' = \bbV\diag(\tbh')\bbV^\top$; i.e., $\bbY' = \bbH'\bbX$. To predict and generate $\bbY'$, we design a learnable parametric function $\hat\bbY' = \Upsilon(\bbX;\bbTheta_\Upsilon)$ via unrolling, similar to SLoG-Net. Estimating both $\bbY'$ and $\tbh'$ from the input signal $\bbX$ is an ill-posed problem, hence we assume the diffused signal $\bbY'$ is close to $\bbX$. Then we consider the following constrained optimization problem to predict (and thus generate) network diffusion outputs,
\begin{align} \label{e:opt_forward_diffusion}
    \min_{\tbh,\bbY'} & \|\bbY' - \bbV\diag(\tbh)\bbV^\top\bbX\|_\textrm{F}^2 + \rho_1\|\bbY'-\bbX\|_\textrm{F}^2\nonumber\\
    &{} \text{s. to }\bar{\bbM}^\top \tbh = \bar{\bbm},  
\end{align}
where $\bar{\bbM}\in\reals^{N\times d},\bar{\bbm}\in\reals^{d}$ will be learned. 
%

Similar to SLoG-Net, we derive ADMM iterations to solve \eqref{e:opt_forward_diffusion} and use the unrolling principle to construct the sub-layers of the network diffusion NN (ND-Net) $\hat\bbY' = \Upsilon(\bbX;\bbTheta_d)$; details are omitted to avoid repetition.
We also consider $K=5$ layers 
and generate predictions as $\hat\bbY'=(\bbX^\top \bbV  \odot \bbV)\tbh[K]=\Upsilon(\bbX;\bbTheta_d)$. Given this architecture, we compose ND-Net $\Upsilon$ and the inverse calibration mapping $\Phi_c^{-1}$ to obtain the pre-training model $\bbY = \Phi_c^{-1}(\Upsilon(\bbX;\bbTheta_\Upsilon);\bbTheta_c)$; see Fig.~\ref{fig:cSLoG-Net} (top). During the pre-training process, the learnable parameters $\{\bbTheta_c, \bbTheta_\Upsilon\}$ are learned from binary data $\ccalT = \{\bbY_q,\bbX_q\}_{q}^Q$. 

\bibliographystyle{IEEEtranS}
%
\bibliography{citations.bib}

\begin{thebibliography}{10}
\providecommand{\url}[1]{#1}
\csname url@samestyle\endcsname
\providecommand{\newblock}{\relax}
\providecommand{\bibinfo}[2]{#2}
\providecommand{\BIBentrySTDinterwordspacing}{\spaceskip=0pt\relax}
\providecommand{\BIBentryALTinterwordstretchfactor}{4}
\providecommand{\BIBentryALTinterwordspacing}{\spaceskip=\fontdimen2\font plus
\BIBentryALTinterwordstretchfactor\fontdimen3\font minus
  \fontdimen4\font\relax}
\providecommand{\BIBforeignlanguage}[2]{{%
\expandafter\ifx\csname l@#1\endcsname\relax
\typeout{** WARNING: IEEEtranS.bst: No hyphenation pattern has been}%
\typeout{** loaded for the language `#1'. Using the pattern for}%
\typeout{** the default language instead.}%
\else
\language=\csname l@#1\endcsname
\fi
#2}}
\providecommand{\BIBdecl}{\relax}
\BIBdecl

\bibitem{ahmed2014blind}
A.~Ahmed, B.~Recht, and J.~Romberg, ``Blind deconvolution using convex
  programming,'' \emph{IEEE Trans. Inf. Theory}, vol.~60, no.~3, pp.
  1711--1732, 2014.

\bibitem{beck2009fista}
A.~Beck and M.~Teboulle, ``A fast iterative shrinkage-thresholding algorithm
  for linear inverse problems,'' \emph{SIAM J. Imaging Sci.}, vol.~2, no.~1,
  pp. 183--202, 2009.

\bibitem{behrmann2019invertible}
J.~Behrmann, W.~Grathwohl, R.~T. Chen, D.~Duvenaud, and J.-H. Jacobsen,
  ``Invertible residual networks,'' in \emph{Proc. Int. Conf. Mach. Learn.},
  2019, pp. 573--582.

\bibitem{Bertsekas_Book_Distr}
D.~P. Bertsekas and J.~N. Tsitsiklis, \emph{Parallel and Distributed
  Computation: Numerical Methods}, 2nd~ed.\hskip 1em plus 0.5em minus
  0.4em\relax Athena-Scientific, 1999.

\bibitem{boyd2011distributed}
S.~Boyd, N.~Parikh, E.~Chu, B.~Peleato, J.~Eckstein \emph{et~al.},
  ``Distributed optimization and statistical learning via the alternating
  direction method of multipliers,'' \emph{Foundations and
  Trends{\textregistered} in Machine learning}, vol.~3, no.~1, pp. 1--122,
  2011.

\bibitem{chen2021graphunroll}
S.~Chen, Y.~C. Eldar, and L.~Zhao, ``Graph unrolling networks: Interpretable
  neural networks for graph signal denoising,'' \emph{IEEE Trans. Signal
  Process.}, vol.~69, pp. 3699--3713, 2021.

\bibitem{DeGrootConsensus}
M.~H. DeGroot, ``Reaching a consensus,'' \emph{J. Am. Stat. Assoc}, vol.~69,
  pp. 118--121, 1974.

\bibitem{feizi2016network}
S.~Feizi, M.~M{\'e}dard, G.~Quon, M.~Kellis, and K.~Duffy, ``Network infusion
  to infer information sources in networks,'' \emph{IEEE Trans. Netw. Sci.
  Eng.}, vol.~6, no.~3, pp. 402--417, 2018.

\bibitem{gama2020spmag}
F.~Gama, E.~Isufi, G.~Leus, and A.~Ribeiro, ``Graphs, convolutions, and neural
  networks: From graph filters to graph neural networks,'' \emph{IEEE Signal
  Process. Mag.}, vol.~37, no.~6, pp. 128--138, 2020.

\bibitem{gama2019tsp}
F.~Gama, A.~G. Marques, G.~Leus, and A.~Ribeiro, ``Convolutional neural network
  architectures for signals supported on graphs,'' \emph{IEEE Trans. Signal
  Process.}, vol.~67, no.~4, pp. 1034--1049, 2019.

\bibitem{gavili2017shift}
A.~Gavili and X.-P. Zhang, ``On the shift operator, graph frequency, and
  optimal filtering in graph signal processing,'' \emph{IEEE Trans. Signal
  Process.}, vol.~65, no.~23, pp. 6303--6318, 2017.

\bibitem{giannakis2016chapter}
G.~B. Giannakis, Q.~Ling, G.~Mateos, I.~D. Schizas, and H.~Zhu, ``Decentralized
  learning for wireless communications and networking,'' in \emph{Splitting
  Methods in Communication, Imaging, Science, and Engineering}, R.~Glowinski,
  S.~J. Osher, and W.~Yin, Eds.\hskip 1em plus 0.5em minus 0.4em\relax
  Springer, 2016, pp. 461--497.

\bibitem{gregor2010lista}
K.~Gregor and Y.~LeCun, ``Learning fast approximations of sparse coding,'' in
  \emph{Proc. Int. Conf. Mach. Learn.}, 2010, p. 399–406.

\bibitem{hogg2012social}
T.~Hogg and K.~Lerman, ``Social dynamics of {D}igg,'' \emph{EPJ Data Sci.},
  vol.~1, pp. 1--26, 2012.

\bibitem{horn2013matrixanalysis}
R.~A. Horn and C.~R. Johnson, \emph{Matrix Analysis}.\hskip 1em plus 0.5em
  minus 0.4em\relax Cambridge University Press, 2013.

\bibitem{hu2016localizing}
C.~Hu, X.~Hua, J.~Ying, P.~M. Thompson, G.~E. Fakhri, and Q.~Li, ``Localizing
  sources of brain disease progression with network diffusion model,''
  \emph{IEEE J. Sel. Topics Signal Process.}, vol.~10, no.~7, pp. 1214--1225,
  2016.

\bibitem{isufi2024gf}
E.~Isufi, F.~Gama, D.~I. Shuman, and S.~Segarra, ``Graph filters for signal
  processing and machine learning on graphs,'' \emph{IEEE Trans. Signal
  Process.}, pp. 1--32, 2024.

\bibitem{kingma2014adam}
D.~P. Kingma and J.~Ba, ``Adam: A method for stochastic optimization,'' in
  \emph{Proc. Int. Conf. Learn. Representations}, 2015, pp. 1--15.

\bibitem{levin2011understanding}
A.~Levin, Y.~Weiss, F.~Durand, and W.~T. Freeman, ``Understanding blind
  deconvolution algorithms,'' \emph{IEEE Trans. Pattern Anal. Mach. Intell.},
  vol.~33, no.~12, pp. 2354--2367, 2011.

\bibitem{li2015unified}
Y.~Li, K.~Lee, and Y.~Bresler, ``Identifiability in bilinear inverse problems
  with applications to subspace or sparsity-constrained blind gain and phase
  calibration,'' \emph{IEEE Trans. Inf. Theory}, vol.~63, no.~2, pp. 822--842,
  Feb 2017.

\bibitem{LingBiConvexCS}
S.~Ling and T.~Strohmer, ``Self-calibration and biconvex compressive sensing,''
  \emph{Inverse Probl.}, vol.~31, no. 115002, pp. 1--31, 2015.

\bibitem{mardani2012admm}
M.~Mardani, G.~Mateos, and G.~B. Giannakis, ``In-network sparsity-regularized
  rank minimization: Algorithms and applications,'' \emph{IEEE Trans. Signal
  Process.}, Mar. 2012 (submitted; see also arXiv preprint arXiv:1203.1570
  [cs.MA]).

\bibitem{monga2021spmag}
V.~Monga, Y.~Li, and Y.~C. Eldar, ``Algorithm unrolling: Interpretable,
  efficient deep learning for signal and image processing,'' \emph{IEEE Signal
  Process. Mag.}, vol.~38, no.~2, pp. 18--44, 2021.

\bibitem{nagahama2022tsp}
M.~Nagahama, K.~Yamada, Y.~Tanaka, S.~H. Chan, and Y.~C. Eldar, ``Graph signal
  restoration using nested deep algorithm unrolling,'' \emph{IEEE Trans. Signal
  Process.}, vol.~70, pp. 3296--3311, 2022.

\bibitem{ortega2018gsp}
A.~Ortega, P.~Frossard, J.~Kovačević, J.~M.~F. Moura, and P.~Vandergheynst,
  ``Graph signal processing: Overview, challenges, and applications,''
  \emph{Proc. IEEE}, vol. 106, no.~5, pp. 808--828, 2018.

\bibitem{pena2016source}
R.~Pena, X.~Bresson, and P.~Vandergheynst, ``Source localization on graphs via
  $\ell_1$ recovery and spectral graph theory,'' in \emph{Proc. IEEE Image,
  Video, and Multidimens. Signal Process. Workshop}, 2016, pp. 1--5.

\bibitem{pinto2012locating}
P.~C. Pinto, P.~Thiran, and M.~Vetterli, ``Locating the source of diffusion in
  large-scale networks,'' \emph{Phys. Rev. Lett.}, vol. 109, no. 068702, pp.
  1--5, 2012.

\bibitem{pu2021learning}
X.~Pu, T.~Cao, X.~Zhang, X.~Dong, and S.~Chen, ``Learning to learn graph
  topologies,'' in \emph{Proc. Adv. Neural. Inf. Process. Syst.}, 2021, pp.
  1--14.

\bibitem{david_blind_sp}
D.~Ram\'{\i}rez, A.~G. Marques, and S.~Segarra, ``Graph-signal reconstruction
  and blind deconvolution for structured inputs,'' \emph{Signal Processing},
  vol. 188, p. 108180, 2021.

\bibitem{DSP_freq_analysis}
A.~Sandryhaila and J.~M.~F. Moura, ``Discrete signal processing on graphs:
  Frequency analysis,'' \emph{IEEE Trans. Signal Process.}, vol.~62, no.~12,
  pp. 3042--3054, June 2014.

\bibitem{sandryhaila2013discrete}
A.~Sandryhaila and J.~M. Moura, ``Discrete signal processing on graphs,''
  \emph{IEEE Trans. Signal Process.}, vol.~61, no.~7, pp. 1644--1656, 2013.

\bibitem{sefer2016diffusion}
E.~Sefer and C.~Kingsford, ``Diffusion archeology for diffusion progression
  history reconstruction,'' in \emph{Proc. IEEE Conf. on Data Mining}, 2014,
  pp. 530--539.

\bibitem{segarra2017blindid}
S.~Segarra, G.~Mateos, A.~G. Marques, and A.~Ribeiro, ``Blind identification of
  graph filters,'' \emph{IEEE Trans. Signal Process.}, vol.~65, no.~5, pp.
  1146--1159, 2017.

\bibitem{shrivastava2019glad}
H.~Shrivastava, X.~Chen, B.~Chen, G.~Lan, S.~Aluru, and L.~Song, ``{GLAD}:
  Learning sparse graph recovery,'' in \emph{Proc. Int. Conf. Learn.
  Representations}, 2020, pp. 1--22.

\bibitem{victor2024icassp}
V.~M. Tenorio, S.~Rey, and A.~G. Marques, ``Blind deconvolution of sparse graph
  signals in the presence of perturbations,'' in \emph{Proc. Int. Conf.
  Acoustics, Speech, Signal Process.}, 2024, pp. 9406--9410.

\bibitem{wang2022invertible}
J.~Wang, J.~Jiang, and L.~Zhao, ``An invertible graph diffusion neural network
  for source localization,'' in \emph{Proc. ACM Web Conf.}, 2022, pp.
  1058--1069.

\bibitem{wang2016blind}
L.~Wang and Y.~Chi, ``Blind deconvolution from multiple sparse inputs,''
  \emph{IEEE Signal Process. Lett.}, vol.~23, no.~10, pp. 1384--1388, 2016.

\bibitem{wasserman2024bnn}
M.~Wasserman and G.~Mateos, ``Graph structure learning with interpretable
  {B}ayesian neural networks,'' \emph{Trans. Mach. Learn. Res.}, pp. 1--27,
  2024.

\bibitem{max2023gdn}
M.~Wasserman, S.~Sihag, G.~Mateos, and A.~Ribeiro, ``Learning graph structure
  from convolutional mixtures,'' \emph{Trans. Mach. Learn. Res.}, pp. 1--29,
  2023.

\bibitem{yang2020admmunroll}
Y.~Yang, J.~Sun, H.~Li, and Z.~Xu, ``{ADMM-CSNet}: {A} deep learning approach
  for image compressive sensing,'' \emph{IEEE Trans. Pattern Anal. Mach.
  Intell.}, vol.~42, no.~3, pp. 521--538, 2020.

\bibitem{chang2024perturbation}
C.~Ye and G.~Mateos, ``Blind deconvolution of graph signals: Robustness to
  graph perturbations,'' \emph{IEEE Signal Process. Lett.}, Dec. 2024
  (submitted; see also arXiv preprint arXiv:2412.15133 [eess.SP]).

\bibitem{chang2024exact}
------, ``Blind deconvolution on graphs: Exact and stable recovery,''
  \emph{Signal Process.}, vol. 230, p. 109864, May 2025.

\bibitem{chang2022eusipco}
------, ``Learning to identify sources of network diffusion,'' in \emph{Proc.
  of European Signal Process. Conf.}, 2022, pp. 727--731.

\bibitem{chang2018eusipco}
C.~Ye, R.~Shafipour, and G.~Mateos, ``Blind identification of invertible graph
  filters with multiple sparse inputs,'' in \emph{Proc. of European Signal
  Process. Conf.}, 2018, pp. 121--125.

\bibitem{zhang2016one}
H.~Zhang, M.~Yan, and W.~Yin, ``One condition for solution uniqueness and
  robustness of both l1-synthesis and l1-analysis minimizations,'' \emph{Adv.
  Comput. Math.}, vol.~42, no.~6, pp. 1381--1399, 2016.

\bibitem{zhang2016towards}
P.~Zhang, J.~He, G.~Long, G.~Huang, and C.~Zhang, ``Towards anomalous diffusion
  sources detection in a large network,'' \emph{ACM T. Internet Techn.},
  vol.~16, no.~1, pp. 1--24, 2016.

\end{thebibliography}

\end{document}